\documentclass[usenatbib]{mn2e}
\usepackage{longtable}
\usepackage{graphicx}
\usepackage{amsmath}
\usepackage{amssymb}
\usepackage{times}
\usepackage{mathtools}
\usepackage{morefloats}
\usepackage{etex}
\pdfoutput=1

\title[G2C2 II: CMRs for Galactic GCs]{G2C2 II: Integrated colour-metallicity relations for Galactic Globular Clusters in SDSS passbands}  
\author[J. Vanderbeke et al.]
{Joachim Vanderbeke$^{1,2}$\thanks{E-mail: Joachimvanderbeke@gmail.com}, Michael
  J. West$^{2,3}$, Roberto De Propris$^4$,  Eric W. Peng$^{5,6}$ , \newauthor
  John P. Blakeslee$^{8,9}$, Andr\'{e}s Jord\'{a}n$^{7}$, Patrick
  C\^{o}t\'{e}$^{8}$, Michael Gregg$^{10,11}$, \newauthor Laura
  Ferrarese$^{8}$, Marianne Takamiya$^{12}$, Maarten Baes$^{1}$ \\ \\
$^{1}$ Sterrenkundig Observatorium, Universiteit Gent, Krijgslaan 281 S9, B-9000 Gent, Belgium\\
 $^{2}$ European Southern Observatory, Alonso de C\'{o}rdova 3107,
 Vitacura, Santiago, Chile\\ 
$^{3}$ Maria Mitchell Observatory, 4 Vestal Street, Nantucket, MA
02554, USA \\
$^{4}$ Finnish Centre for Astronomy with ESO (FINCA), University of Turku, V{\"a}is{\"a}l{\"a}ntie 20,
   FI-21500 Piikki{\"o}, Finland \\
$^{5}$ Department of Astronomy, Peking University, Beijing 100871, China \\
$^{6}$ Kavli Institute for Astronomy and Astrophysics, Beijing 100871,
China \\
$^{7}$ Instituto de Astrof\'isica, Facultad de F\'isica,
Pontificia Universidad
  Cat\'olica de Chile, Av.\ Vicu\~na Mackenna
    4860, 7820436 Macul, Santiago, Chile \\
$^{8}$ Herzberg Insititute of Astrophysics, National Research Council,
Victoria, BC V9E2E7, Canada \\
$^{9}$ Department of Physics and Astronomy,
Washington State University, 1245 Webster Hall, Pullman, WA
99163-2814, USA \\
$^{10}$ Department of Physics, University of California, Davis, CA 956160,
USA\\
$^{11}$ Institute for Geophysics and Planetary Physics, Lawrence
Livermore National Laboratory, L-413, Livermore, CA 94550, USA \\
$^{12}$ Physics and Astronomy Department, University of Hawaii Hilo, Hilo, HI
96720, USA }

\newcommand{\degree}{\ensuremath{^\circ}}

\begin{document}

\date{Accepted. Received }

\maketitle

\label{firstpage}

\begin{abstract}
We use our integrated SDSS photometry for 96 globular clusters in $g$ and $z$, as well as $r$ and $i$ photometry for a subset of
56 clusters, to derive  the integrated colour-metallicity relation (CMR) for Galactic globular clusters. We compare this relation to 
previous work, including extragalactic clusters, and examine the influence of age, present-day mass function variations, structural parameters 
and the morphology of the horizontal branch on the relation. Moreover, we scrutinise the scatter introduced 
by foreground extinction (including differential reddening) and show that the scatter in the colour-metallicity 
relation can be significantly reduced combining two reddening laws from the literature. In all CMRs we find some
low-reddening young GCs that are offset to the CMR. Most of these outliers are associated with the Sagittarius system. 
Simulations show that this is due less to age than to a different enrichment history. Finally, we introduce colour-metallicity 
relations based on the infrared Calcium triplet, which are clearly non-linear when compared to $(g^\prime-i^\prime)$ 
and $(g^\prime-z^\prime)$ colours.
\end{abstract}


\section{Introduction}

Globular clusters (hereafter GCs) form during the earliest stages of galaxy formation. In nearly all galaxies that have been studied, GCs 
follow a nearly universal luminosity function \citep[e.g.,][]{Rejkuba2012}, exhibit a bimodal colour distribution (e.g., \citealt{Zepf1993, 
Ostrov1993, Whitmore1995, Mieske2006, Peng2006, Strader2006, Mieske2010, Faifer2011, Yoon2011b} and references therein) and a
strong correlation between the number of globular clusters and the luminosity of their parent galaxies (specific frequency --\citealt{Harris1981}). 
These all imply that the formation of globular clusters has been intimately related to the early assembly history of galaxies \citep{Harris1991}.
Therefore, the properties of globular clusters allow us to use them as fossil tracers of the initial stages of galaxy formation and evolution (see \citealt{West2004} and \citealt{Brodie2006} for reviews).

Although most globular clusters show evidence of extended star formation and enrichment histories (see review by \citealt{Gratton2012}), the 
vast majority are nevertheless almost homogeneous in iron \citep{Marino2013} and have large ages ($> 10$ Gyr, \citealt{Chaboyer1998,
Strader2005,Puzia2005}), making them still the best available approximation of single stellar populations. Their integrated colours are largely 
dominated by light from K-giants and this allows for a more straightforward interpretation (in terms of age, metallicity, etc.) than the more 
complex stellar populations of galaxies. 

The bimodal distribution of GC colours corresponds, at least in our Galaxy, to a bimodal distribution of metallicities 
\citep[e.g.][]{Zinn1985}, with blue clusters being more metal poor than their red counterparts. Moreover, these two GC
populations are also kinematically distinct \citep{Sharples1998, Zepf2000, Cote2001, Cote2003, Peng2004, Strader2011,
Pota2013}. However, it is still uncertain whether the bimodal distributions of GC colours observed in more distant galaxies
can be generalized into bimodal {\it metallicity} distributions \citep{Cantiello2007,Galleti2009, Foster2010, Foster2011, 
AlvesBrito2011, Caldwell2011, ChiesSantos2011a,ChiesSantos2011b,Usher2012}. There is no colour that yields a univocal
mapping to metallicity. \cite{Yoon2006} and \cite{Richtler2006}  showed that a non-linear relation between metallicity and
integrated colour can transform  a unimodal metallicity distribution into a bimodal colour distribution. Vice versa,
\cite{Yoon2011a} demonstrated that the bimodal colour distributions could be transformed into metallicity distributions
consisting of a sharp peak with a metal-poor tail, similar to the metallicity distribution functions of resolved field stars in 
nearby elliptical galaxies \citep[e.g.][]{Harris2002}.

Colour-metallicity relations (CMR), calibrated to objects of known metal abundance are essential to correctly interpret the 
colour distributions in external galaxies. The first empirical relationships between colour and [Fe/H] were approximately 
linear \citep{Brodie1990,Couture1990, KisslerPatig1997, KisslerPatig1998}, while more recent studies tend to prefer 
non-linear CMRs \citep{Peng2006,Blakeslee2010,Sinnott2010,Usher2012}. \cite{Faifer2011} find a linear CMR, but their 
sample was lacking in metal-poor GCs, which are the clusters invoking the non-linearity of the CMRs.

Some stellar population models also predict a non-linear CMR \citep{Yoon2006,Yoon2011a,Yoon2011b,Cantiello2007} 
and studies based on optical-NIR colours \citep{Chiessantos2012,Blakeslee2012} provide further evidence that the 
non-linearity of the CMR is unavoidable even in the presence of unimodal metallicity distributions. Moreover, the 
metallicity distributions seem usually less bimodal than the optical colour distributions, which suggests that at least part 
of the observed colour bimodality is caused by the nonlinearity of the CMR \citep{Blakeslee2012}. Although most of these
studies agree on the non-linear form of the CMR, further data sets are indispensable to extend the metallicity range, to 
better calibrate the CMR and to capture its details.

In Vanderbeke et al. (2013, in press, hereafter Paper I) we have presented integrated photometry for 96 globular clusters in SDSS 
passbands $g^\prime$ and $z^\prime$, as well as $r^\prime$ and $i^\prime$ photometry for a subset of 56 clusters. 
This paper discusses the dataset and the error statistics of the sample. Here we use these data to construct the Galactic 
colour-metallicity relation. We discuss the CMRs for the different colour combinations in Section~\ref{sec:cmr} and study
 the influence of (differential) reddening, horizontal branch (HB) morphology, age, present-day mass function variations, structural parameters and 
contamination correction on the scatter in the CMR. In Section~\ref{sec:colourbim} we investigate the colour bimodality 
of our sample, and summarize the results in Section~\ref{sec:results}. 

\section{The colour-metallicity relation for Globular Clusters} \label{sec:cmr}

We use the photometry described in detail in Paper I to refine the colour-metallicity relations for GCs. For 
some clusters, both SDSS and CTIO magnitudes were obtained. Motivated by the findings of Paper I, we use CTIO magnitudes 
in all cases, except for Pal~3 and Pal~13. For Galactic clusters, metallicities are taken from the compilation of 
\cite{Harris1996},  using the latest version, and are based on the \cite{Carretta2009}  scale supplemented with 
data from \cite{Armandroff1988}. However, our Galaxy lacks the high metallicity (solar or more) GCs that are 
encountered in bright cluster ellipticals; following \cite{Peng2006} and \cite{Blakeslee2010} we will supplement 
our data with photometry (and metallicities) for globular clusters in M 49 and M 87, taken from the SDSS and the 
ACS Virgo  Cluster Survey \citep{Cote2004} in Section \ref{sec:EGcmr}. There is also a lack of very metal-poor clusters  
in both the Galactic and extragalactic samples; some globular clusters in dwarf spheroidals appear to be more 
metal-poor than the most extreme such objects in our Galaxy.

\begin{figure}
\centering 
\includegraphics[scale=0.87,trim=2.8cm 13.1cm 3cm 6cm] {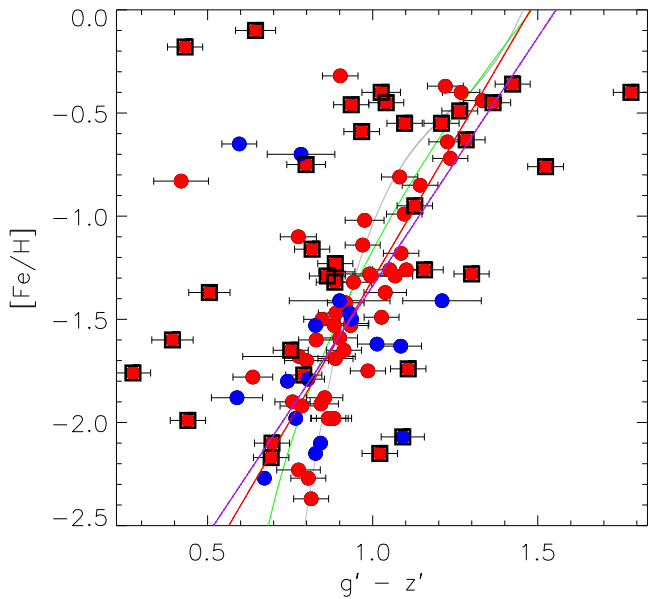}
\caption{[Fe/H] as a function of $g^\prime - z^\prime$ for 96 Galactic
  GCs. Red circles represent clusters with CTIO data, blue circles are
  used for clusters with SDSS data. Clusters with $E(B-V) \geq 0.35$ are indicated
  with boxes and are excluded to make the fits. The purple line is a
  linear robust fit to all the data, while the red line is fitting only the
  CTIO data. As a reference, CMRs from the literature
  \citep{Sinnott2010,Blakeslee2010} are
  also  presented as green and grey lines (resp.). The four metal-rich low-reddening GCs that are offset the CMRs are E~3 and Terzan~7 (CTIO data) and Whiting~1 and Pal~1 (SDSS data). See text for more
details. }  
\label{GalgzCMR}
\end{figure}

In Fig.~\ref{GalgzCMR} we present [Fe/H] as a function of $g^\prime - z^\prime$ for 96 Galactic GCs. Red 
filled circles represent CTIO photometry, blue filled circles are used for clusters with SDSS data. The 
previous CMRs from \cite{Blakeslee2010} and \cite{Sinnott2010} are shown as grey and green lines.  For completeness, 
we show all clusters in Fig.~\ref{GalgzCMR}, but only the clusters with extinction $E(B-V) < 0.35$ \citep{Harris1996}
are used to fit the CMR. 

We fit a straight line to the data using the method of least absolute deviation (robust fitting) as this is
less sensitive to outliers \citep{Armstrong1978}. The best robust fit to the CTIO data only is given by:
\begin{equation} \label{eq:Galgz_ctio_only}
[\text{Fe/H}]= (-4.04\pm0.04) + (2.74\pm0.04)\times (g^\prime - z^\prime)
\end{equation}

and is represented by a red line in the figure, while the best fit to both SDSS and CTIO data is given by:
\begin{equation} 
[\text{Fe/H}]= (-3.75\pm0.04) + (2.41\pm0.05)\times (g^\prime - z^\prime),
\end{equation}

which is the purple line in Fig.~\ref{GalgzCMR}. The errors on the coefficients are computed by a bootstrap 
method. We note that the CMR of \cite{Blakeslee2010} seems to overestimate [Fe/H] for the relatively more metal-rich 
clusters, while our linear relations do not fit well the metal-poor end of the metallicity range. There is considerable
scatter around the relation, more than would be expected simply from photometric errors. This appears to be related
to foreground reddening and is discussed in detail in Section \ref{sec:reddening}.

SDSS data for low-reddening clusters show more scatter around the existing relations than the corresponding CTIO data. 
When computing the horizontal RMS for these two subsamples with respect to the CMR of \cite{Blakeslee2010}, we find
a RMS of 0.14 for the low-reddening CTIO subsample, but a RMS of 0.20 for the low-reddening SDSS subsample. 
This might be due to the saturation issues for SDSS data discussed in Paper I. On the metal-rich
side, four low-reddening clusters are offset from the relation. These clusters are E~3 and Terzan~7 (CTIO data) and 
Whiting~1 and Pal~1 (SDSS data). In Paper I we raised some sky determination issues which affect the obtained magnitudes of E~3 and Terzan~7. For the position in colour-metallicity space of these and other clusters, we refer to 
Fig.~\ref{EGgzCMR}. These and other outliers are discussed in Section~\ref{sec:cmroutliers}. 

Due to the sizable scatter, the limited metallicity range and the low sample size, it is not justifiable to fit a higher-order
polynomial to the Galactic data. The linear fit is a reasonable approximation for the Galactic data only, when considering 
the limitations of the sample. The fit however is not satisfactory, thus we will address this issue again in 
Section~\ref{sec:EGcmr}, where we will include extragalactic data from the literature to extend the metallicity range and 
the sample size.  

\begin{figure}
\centering 
\includegraphics[scale=0.87,trim=2.8cm 13.1cm 3cm 6cm] {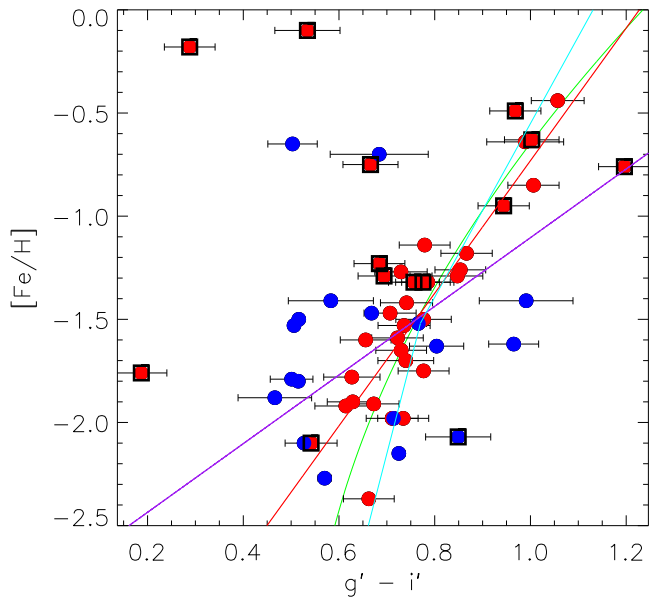}
\caption{[Fe/H] as a function of $g^\prime - i^\prime$ for 56 Galactic
  GCs. Legend as in Fig.~\ref{GalgzCMR}. The cyan line is
  another CMR from the literature \citep{Usher2012}. The two low-reddening metal-rich GCs based on SDSS data are Whiting~1 and Pal~1. The SDSS outliers at $[Fe/H]\sim-1.5$ with small photometric uncertainties are NGC~5272 and NGC~6205. }  
\label{GalgiCMR}
\end{figure}

Fig.~\ref{GalgiCMR} shows the CMR for the $g^\prime - i^\prime$ colour. As a reference, the CMRs published in 
\cite{Sinnott2010} and \cite{Usher2012} are presented as green and cyan lines. \cite{Sinnott2010} presented a 
$g^\prime - i^\prime$ CMR for clusters in NGC~5128, using Milky Way clusters to convert their  [MgFe]$^\prime$
index to [Fe/H], but their CMR  has not been compared to $g^\prime - i^\prime$ colours for Galactic GCs. It is 
generally assumed that Galactic GCs are not intrinsically different from extragalactic GCs \citep[e.g.][]{Foster2010}, 
although \cite{Usher2012} note that differences in the CMR could be driven by differences in the age or the IMF of 
GCs between galaxies. These possibilities will be further discussed in Sections~\ref{sec:age} and \ref{sec:IMF}.  

Galactic GC colours based on CTIO data compare well to the extragalactic CMR of \cite{Sinnott2010}. Only clusters 
suffering severe reddening are outliers. The two low-reddening metal-rich GCs based on SDSS data are 
Whiting~1 (associated to the Sagittarius system) and Pal~1. For E~3 and Terzan~7, discussed above, 
no $i^\prime$-band observations were performed. Nevertheless, it is clear again that the low-reddening SDSS 
data has more scatter around the \cite{Sinnott2010} CMR than the low-reddening CTIO data. NGC~5272 and 
NGC~6205 are the low-reddening outliers at $[Fe/H]\sim-1.5$ with small photometric uncertainties. This reinforces the 
suspicion that the SDSS magnitudes of bright clusters are affected by saturation of their bright stars (as discussed 
in Paper I).  We will discuss this further in a future paper dealing with the colour-magnitude diagrams.

The best robust fit for the CTIO data only is given by
\begin{equation} \label{eq:Galgi_ctio_only}
[\text{Fe/H}]=  (-3.94\pm0.05) + (3.21\pm0.06)\times (g^\prime - i^\prime),
\end{equation}while including the SDSS data results in 
\begin{equation} 
[\text{Fe/H}]=  (-2.77\pm0.02) + (1.66\pm0.03)\times (g^\prime - i^\prime).
\end{equation} 

Errors on this relation were determined using a bootstrap method. It is clear that the relation based on CTIO 
data only compares well to the extragalactic relation published by \cite{Sinnott2010}, while the scatter in the SDSS data 
results in an inconsistent fit when using all available data. The relative lack of metal-poor clusters makes it hard to
constrain the metal-poor part of the CMR. 

\cite{Usher2012} proposed a broken line fit as their [Z/H]-$(g -i)$ relation (their Eq.~10) and give a conversion between
[Z/H] and [Fe/H] (their Eq.~1). Plotting this relation as a cyan line on Fig.~\ref{GalgiCMR} we find that their relation predicts
slightly redder colours for metal-poor clusters than our observed $g^\prime - i^\prime$ colours. Nevertheless, our sample 
has only a couple of metal-poor objects with available $g^\prime - i^\prime$ colours. 

\begin{figure}
\centering 
\includegraphics[scale=0.87,trim=2.8cm 13.1cm 3cm 6cm] {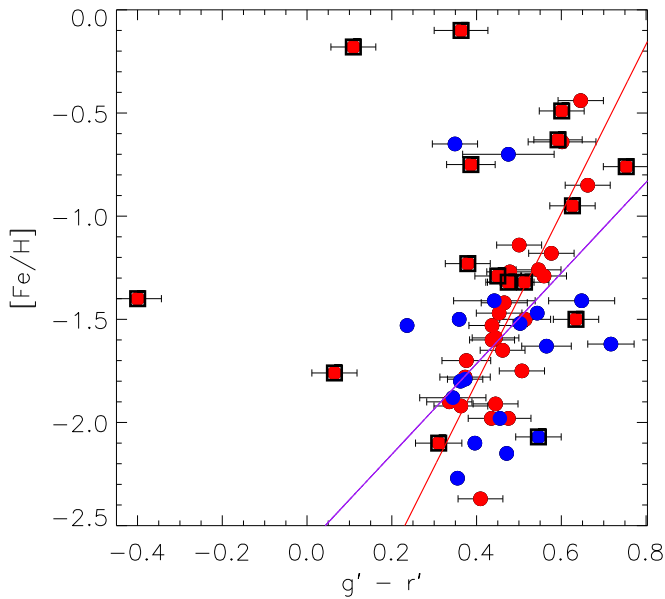}
\caption{[Fe/H] as a function of $g^\prime - r^\prime$ for 58 Galactic
  GCs.  Legend as in Fig.~\ref{GalgzCMR}. The two low-reddening metal-rich GCs based on SDSS data are Whiting~1 and Pal~1. The fitted relations are given by Eqs.~\ref{eq:Galgr_ctio_only} and \ref{eq:Galgr_ctio_and_SDSS}. }  
\label{GalgrCMR}
\end{figure}

Fig.~\ref{GalgrCMR} presents [Fe/H] as a function of $g^\prime - r^\prime$ for 58 GCs. Again, the more significant
outliers are affected by high foreground reddening, and, the scatter for low-reddening GCs in the SDSS data is significantly
larger than for the CTIO data. When robustly fitting the low-reddening CTIO data, we find:
\begin{equation} \label{eq:Galgr_ctio_only}
[\text{Fe/H}]=  (-3.44\pm0.06) + (4.10\pm0.12)\times (g^\prime - r^\prime),
\end{equation} while including the SDSS data results in 
\begin{equation} \label{eq:Galgr_ctio_and_SDSS}
[\text{Fe/H}]=  (-2.59\pm0.03) + (2.20\pm0.06)\times (g^\prime - r^\prime).
\end{equation}

In this colour the CMR is even closer to being linear, which is expected when considering the limited wavelength baseline 
of $g^\prime-r^\prime$ and hence its relatively weak sensitivity to metal abundance.

\subsection{Outliers in the CMR}\label{sec:cmroutliers}

We here discuss briefly the properties of some of the most significant low-reddening outliers from the Galactic CMR. 

As discussed in Paper I, E~3 and Terzan~7 are very poor clusters and were both observed with the CTIO 0.9~m telescope during conditions with the sky having a higher surface brightness than the average cluster surface brightness (sky-subtracted, within the half-light radius). As a consequence, small variations in the sky determination can significantly affect the obtained magnitudes. Moreover, E~3 suffers from considerable extinction along the line of sight ($E(B-V)\sim0.3$).  

Pal~4, a GC initially suspected to be a dwarf galaxy, is one of the only low-reddening globular clusters which has a $g-z$ colour redder than expected based on its metallicity. However, this offset can be partly explained by its large colour uncertainty ($\sigma_{g-z}\sim0.12$).

Both Whiting~1 and Pal~1 are faint clusters, resulting in very poor CMDs. For the sake of completeness, we provide more details on the photometric analysis performed in Paper I for the latter clusters. For Whiting~1, one candidate non-member star was selected in the CMD but had no proper motions, so the star was not removed. Pal~1 is located at a Galactic latitude of $19.03\degree$ (resulting in a foreground reddening of $E(B-V)\sim0.15$), so some foreground stars are expected. Based on the CMD, seven candidate outliers were selected in the same colour-magnitude region, five of which had known proper motions in the NOMAD catalog \citep{Zacharias2005}, including the four brightest candidates. We decided to remove all seven candidate outliers (because all candidates are in the same CMD region), which resulted in magnitude corrections as presented in Paper I. Not performing the magnitude corrections would result in $g-z=0.82$, thus moving the cluster towards the CMR (to a position close to Whiting~1).  

Despite the photometric uncertainties, it is interesting to discuss these clusters in some more detail. In their study of the Galactic outer halo, \cite{vandenbergh2004} found several similarities for the properties of Pal~1 and Terzan~7 (e.g. $r_h<7pc, [Fe/H]>-0.7$), suggesting that these clusters have similar formation and evolutionary histories. These authors explain the existence of such metal-rich GCs in the outer Galactic halo by a formation in dwarf spheroidal galaxies. This appears probable for Terzan~7, a cluster often associated with the Sagittarius system (e.g. \citealt{Geisler2007}). Moreover, these clusters have estimated ages lower than 8 Gyrs (\citealt{Rosenberg1998,Geisler2007}). Hence, it is possible that these clusters are outliers on the CMR, because they have a different formation history. \cite{Forbes2010} associated Pal~1 as a probable member of the Canis Major dwarf and confirm the similarities between Pal~1 and the Sagittarius dSph GCs in their age-metallicity relation. Nevertheless, other GCs associated by \cite{Forbes2010} to the Canis Major system (including NGC~1851, NGC~1904, NGC~2298, NGC~2808, NGC4590 and Rup~106) are very close to the CMR of \cite{Blakeslee2010}.

Whiting~1 is another young ($\sim6.5$ Gyr) GC associated with the Sagittarius dwarf spheroidal galaxy, hence another object that originated in a dwarf galaxy that has since been disrupted by the tidal forces of the Milky Way \citep{Carraro2005,Carraro2007}. This intermediate-metallicity object is clearly offset of the CMR presented in Fig.~\ref{GalgzCMR}. 

\cite{Lotz2004} find that dE GC candidates are as blue as the metal-poor GCs of the Milky Way. In the discussion, these authors assume that their sample of dE GCs is dominated by old and metal-poor GCs. However, in the above analysis, Pal~1, Whiting~1 and Terzan~7 are younger GCs at intermediate metallicities, associated with dSph galaxies. This again stresses the peculiarity of these objects.

In Sections~\ref{sec:age} to \ref{sec:chemistry} we will reevaluate the position in colour-metallicity space of these clusters. 

\subsection{Calcium Triplet metallicity scale}

Though gaining importance and attention, the infrared Calcium Triplet (CaT) is not yet generally accepted as a metallicity
indicator \citep[e.g.][and references therein]{Foster2010,Usher2012}. Nevertheless, homogeneous CaT measurements 
from \cite{Saviane2012} allow us to produce a CMR based on our $g^\prime r^\prime i^\prime z^\prime$ colours and
metallicity based on this indicator. 

\begin{figure*}
\centering 
\includegraphics[scale=0.95,trim= 3cm 13.2cm 0cm 8cm]{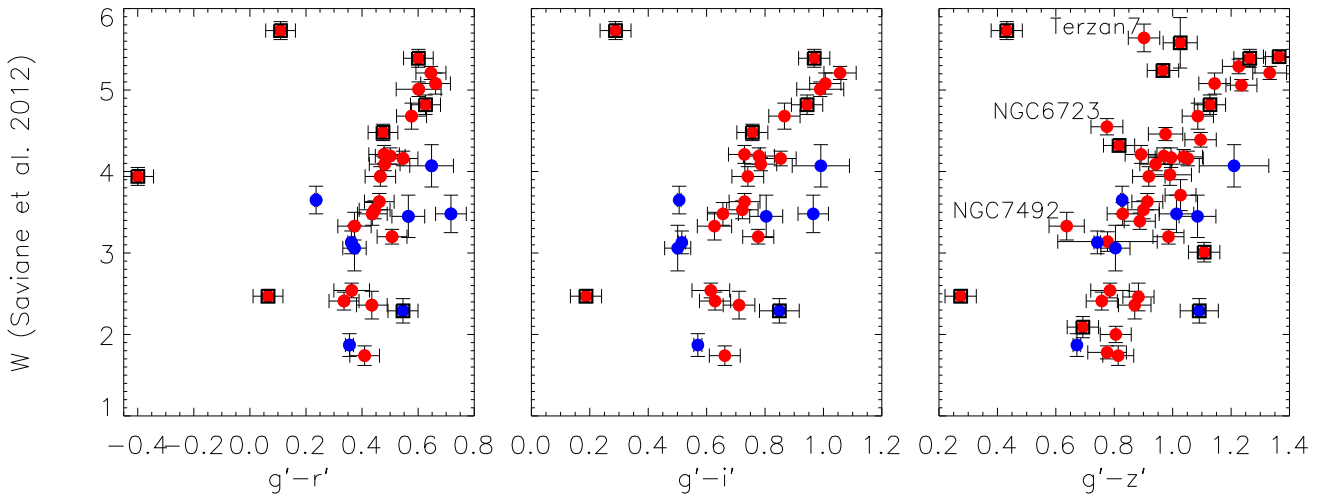}
\caption{Colour-metallicity relations for CaT $W^\prime$ and $g^\prime r^\prime i^\prime z^\prime$
  colours. Legend as in Fig.~\ref{GalgzCMR}. Black boxes
  indicate clusters with $E(B-V)\geq0.35$. It is clear the CaT
  metallicity indicator behaves non-linear when compared to the $g^\prime
- i^\prime$ and $g^\prime- z^\prime$ colours. }
\label{Wsaviane}
\end{figure*}

In this section we compare our colours with $W^\prime$, which is the sum of the equivalent widths of the two strongest 
CaT lines ($\lambda8542 \text{, } \lambda8662$) corrected for the HB level ($W^\prime=W_{8542}+W_{8662}-a(V-V_{HB})$,
\citealt{Armandroff1991}). Fig.~\ref{Wsaviane} presents the CMR for the CaT $W^\prime$ parameter as a function of the
$g^\prime r^\prime i^\prime z^\prime$ colours. Some of the findings of the previous section are confirmed: the scatter 
for the SDSS data is larger than for the CTIO data and for the CTIO data, the scatter is closely related to the reddening 
estimate. Although the metallicity range is limited, the figure suggests that $W^\prime$ is non-linear with $g^\prime-i^\prime$ 
and $g^\prime-z^\prime$. Even for $g^\prime-r^\prime$ the relation seems slightly non-linear. Nevertheless, 
it is again clear that this latter colour has a relatively weak sensitivity to metal abundance, as a consequence of its limited
wavelength baseline. 

Based on the transmission curves, one would expect that $g^\prime - z^\prime$ is the colour most sensitve to $W^\prime$,
because both $\lambda8542$ and $\lambda8662$ fall within $z^\prime$. Indeed, for the clusters with available photometry,
$g^\prime -z^\prime$ shows the largest dynamic range.

Terzan~7, one of the GCs associated to the Sagittarius system, is the metal-rich low-reddening cluster
which is the outlier with respect to the general relation in $g^\prime
- z^\prime$. Note that Terzan~7 was also an outlier in the $(g^\prime- z^\prime)\text{-[Fe/H]}$ CMR presented in Fig.~\ref{GalgzCMR}.

\subsection{The colour-metallicity relation including Extragalactic Globular Clusters} \label{sec:EGcmr}
The Milky Way lacks both very metal-poor and very-metal rich GCs. Here we address the issues raised in the previous 
sections by including extragalactic GCs from the literature. \cite{Peng2006} presented colours for GCs in the giant 
ellipticals M~49 and M~87, for which spectroscopic metallicities were published in \cite{Cohen1998,Cohen2003}. These 
clusters are added to our Galactic sample; they provide extra leverage especially at the metal-rich end of the relation.

\begin{figure*}
\centering 
\includegraphics[scale=0.9,trim=2.8cm 13.1cm 3cm 1cm] {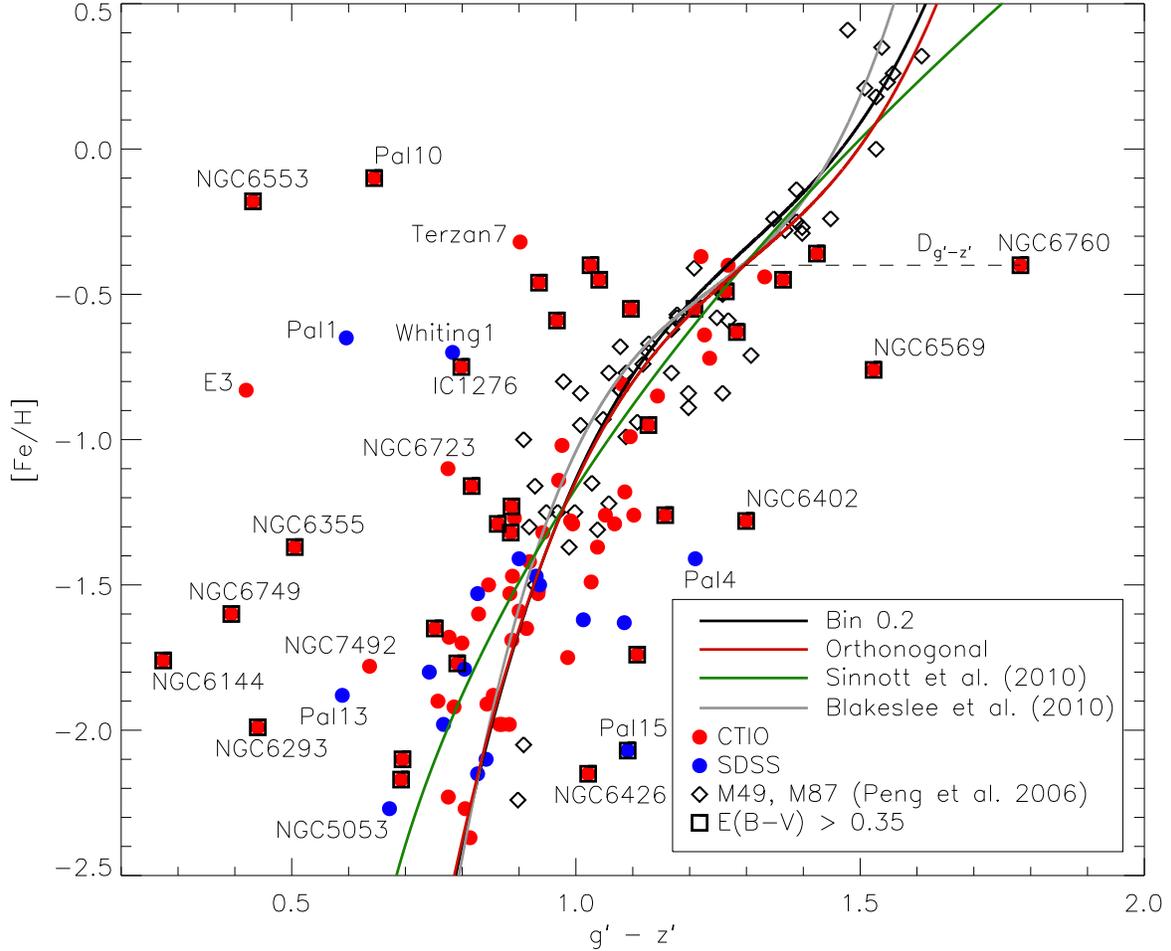}
\caption{[Fe/H] as a function of $g^\prime - z^\prime$ for our sample of Galactic GCs
  and data of extragalactic GCs from the literature. Clusters from our CTIO and SDSS samples are represented with filled circles, while
literature data of M~49 and M~87 \citep{Peng2006} is represented by
diamonds. The black line presents the CMR derived for all low-reddening
GCs, fitting colours binned by metallicity bins of 0.2 (as given by Eq.~\ref{gzbinned}). The red line
is the final CMR (given by Eq.~\ref{eq:gzPython}) obtained by minimizing the orthogonal distance. The grey and
green lines present CMRs from the literature
\citep{Blakeslee2010,Sinnott2010}. The black dashed line presents the distance $D_{g^\prime-z^\prime}$, as defined in Eq. \ref{eq:D}. See text for more details. }
\label{EGgzCMR}
\end{figure*}

We present all available data in  Fig.~\ref{EGgzCMR}. Several Galactic GCs suffer from high reddening ($E(B-V)\geq0.35$) 
and are indicated with boxes in the figure. For clarity, no error bars are presented in this figure. We also show, in the
same figure, CMRs from the literature.

In the previous sections it became clear that most of the scatter in the CTIO data is caused by the uncertainty 
in the reddening correction, which will be scrutinised in Section~\ref{sec:reddening} (while for SDSS data saturation may also play a role). Therefore we exclude clusters 
with high reddening to make the fits. Because the scatter in colour is still significant, we binned the data points in bins 
of 0.2 dex in metallicity. We then computed the median colour for each bin, obtaining a representative colour for each
metallicity bin and fitted a cubic polynomial to the binned data. This resulted in a CMR:
\begin{multline}\label{gzbinned}
[\text{Fe/H}] =  -22.69 \pm   1.54 +  ( 47.81  \pm4.02)  \times (g^\prime
-z^\prime ) \\  -(35.27\pm 3.44) \times(g^\prime -z^\prime )^2   + ( 9.01 \pm 0.96)  \times(g^\prime -z^\prime )^3   
\end{multline}
which is plotted as a black solid line in Fig.~\ref{EGgzCMR}. Note that the uncertainties on the coefficients are large 
because of the small number of degrees of freedom and because we use the bin size as the metallicity uncertainty in the bootstrapping routine. Using the homogeneous \cite{Saviane2012} [Fe/H] values where 
possible does not significantly change the fitted relation. Compared to \cite{Blakeslee2010} we find good agreement 
over the metallicity range considered. When comparing our CMR to \cite{Sinnott2010}, there are larger differences on 
both the metal-poor and metal-rich end. This is partly due to the fact that \cite{Sinnott2010} used a different metallicity 
range for their fit ($-2.2 < $ [Fe/H] $< -0.5$). Nevertheless the two CMRs compare reasonably well up to [Fe/H]$\sim0$. 
On the blue end of the relation, the difference is somewhat larger, stressing the importance of obtaining additional data 
at the low metallicity end. 

As has been discussed in Section~\ref{sec:cmroutliers}, some clusters are strong outliers: Whiting~1, Terzan~7 (clusters associated with the Sagittarius dSph) and Pal~1 (probably related to the Canis Major system). Another outlier is E~3, an old cluster with uncertain photometry. We make a new subset, excluding these clusters from the low-reddening sample. Using 
this subset, we performed an optimization algorithm in \textsc{python}, minimizing the orthogonal distance to the CMR.
This resulted in our final CMR:
\begin{multline}\label{eq:gzPython}
[\text{Fe/H}]=-22.68 +  48.04\times (g^\prime - z^\prime) \\ -35.63 \times (g^\prime -
z^\prime)^2 +    9.12\times(g^\prime -z^\prime)^3, 
\end{multline}
presented as a red line in Fig.~\ref{EGgzCMR}. This relation compares very well to the binned relation given by Eq.~\ref{gzbinned}, with all coefficients within the error bars. Compared to the CMR of \cite{Sinnott2010}, there is a significant offset at both the metal-rich and metal-poor end.  

Visual inspection might suggest that the [Fe/H] range of $-1$ to $-0.5$ reveals a larger spread in $g-z$ for the M~49 and M~87 GCs than for the MW GCs. However, this is misleading, as the overall RMS for the extragalactic GCs is 0.069, while the RMS for the extragalactic GCs with [Fe/H] between $-1$ and $-0.5$ equals 0.074. Note that the RMS for the low-reddening Galactic GCs amounts 0.16. 

\subsection{Colour uncertainties due to Reddening} \label{sec:reddening}

It is clear from Fig.~\ref{GalgzCMR} that most of the CMR outliers, which are not related to dSph galaxies or do not have uncertain photometry, 
suffer from high reddening. Note that not only Galactic studies suffer from this issue: \cite{Kim2013M31} also indicate that 
colours of M~31 GCs are very susceptible of reddening uncertainties. This resulted in colour scatter so large that no 
meaningful comparison could be made with their models.  

\begin{figure}
\centering 
\includegraphics[scale=0.87,trim= 3cm 13cm 0 6cm] {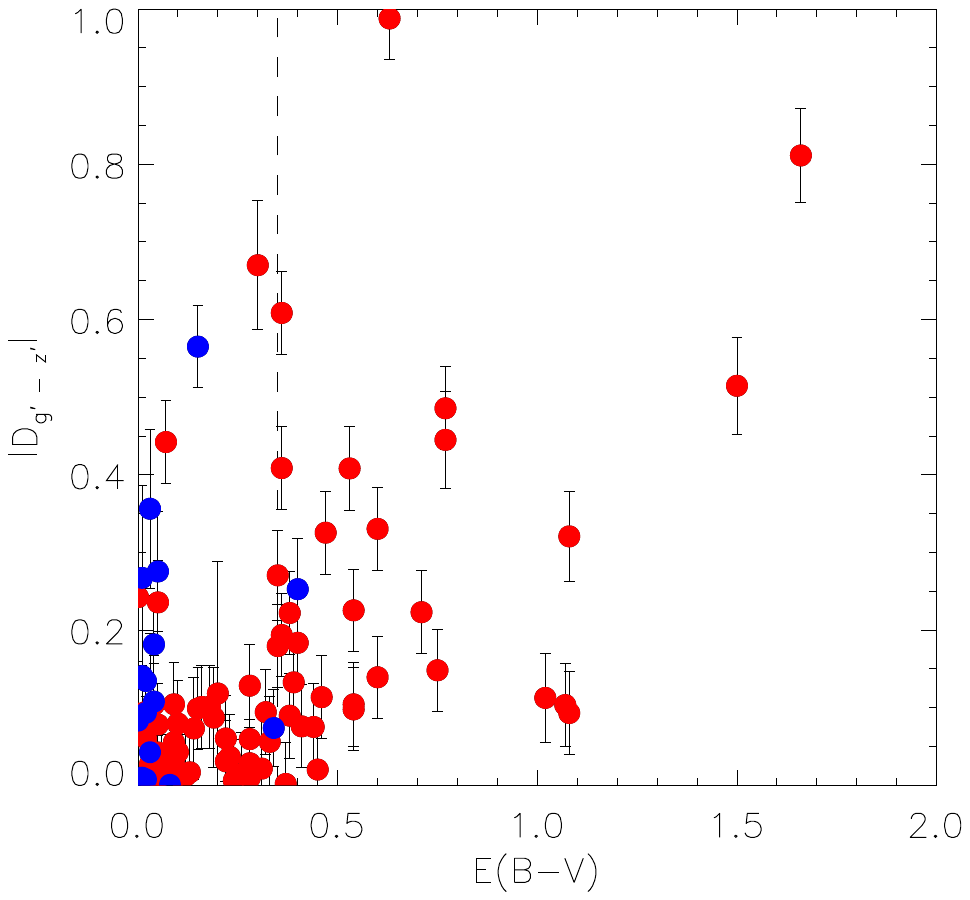}
\caption{Absolute colour difference $|D_{g^\prime - z^\prime}|$ (as defined in Eq.~\ref{eq:D}) as a
  function of $E(B-V)$ \citep{Harris1996}. Legend as in Fig.~\ref{GalgzCMR}. The vertical dashed line
  indicates $E(B-V)=0.35$. }
\label{distance_function_reddening}
\end{figure}

To get a handle on the error introduced by the reddening estimates, we  define a new parameter, which is the colour
difference between the final CMR (Eq.~\ref{eq:gzPython}) assuming the metallicity is accurately known and the observed 
colour: 
\begin{equation} \label{eq:D}
D_{g^\prime - z^\prime} = CMR^{-1}([\text{Fe/H}]) - (g^\prime -
z^\prime)_{observed},
\end{equation}
as indicated in Fig.~\ref{EGgzCMR}. This new parameter is positive (negative) when the observed colour is bluer (redder) 
than the colour predicted by our final CMR. In Fig.~\ref{distance_function_reddening} we plot the distance $|D_{g^\prime -
z^\prime}|$ to the CMR as a function of the reddening $E(B-V)$ (from \citealt{Harris1996} (2010 edition), which is a 
compilation of \citealt{Reed1988,Webbink1985} and \citealt{Zinn1985}). It is clear that a significant part of the scatter is 
caused by the uncertainties in the reddening estimates. Note that the scatter is relatively larger for low-reddening clusters 
with SDSS data than with CTIO data. 

\begin{figure*}
\centering 
\includegraphics[scale=0.87,trim= 3cm 13cm 8cm 6cm]
{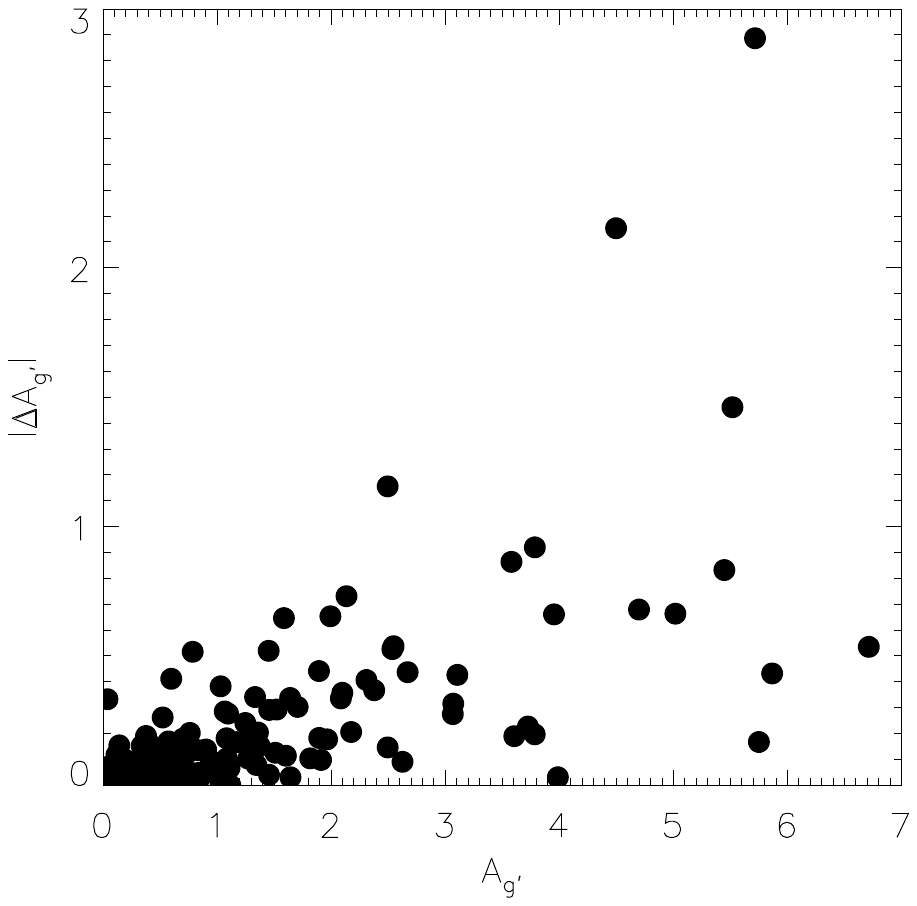}\includegraphics[scale=0.87,trim=
3cm 13cm 8cm 6cm] {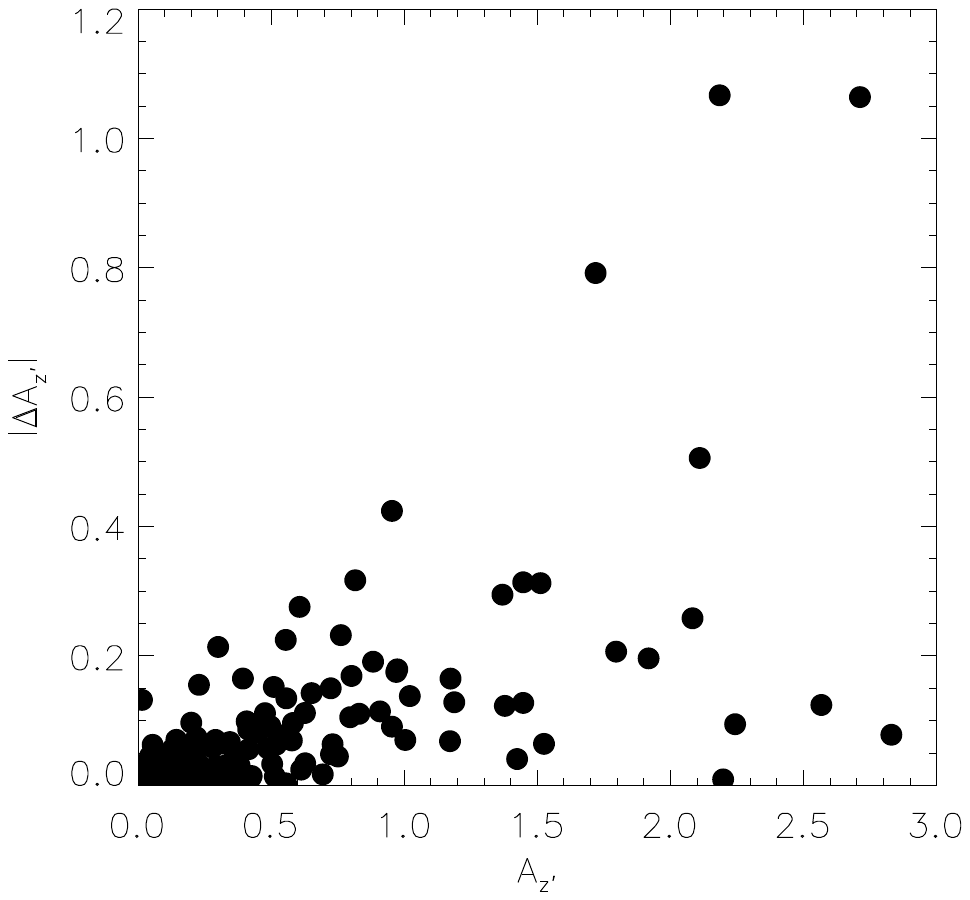}
\caption{Absolute reddening correction difference (defined in
  Eqs.~\ref{eq:deltaAg} and \ref{eq:deltaAz}) as a function of the
reddening coefficient for $g^\prime$ and $z^\prime$ \citep{Schlafly2011}. The uncertainty
on $A_{g^\prime}$ is much larger than on $A_{z^\prime}$.  }
\label{fig:AgAzdiffCardelliSchlafly}
\end{figure*}

To get a better notion of the uncertainty on the reddening estimate, we present in Fig.~\ref{fig:AgAzdiffCardelliSchlafly} 
the absolute difference between the reddening corrections based on \cite{Cardelli1989} (C89) and \cite{Schlafly2011} (S11, which was used in Paper I) 
with:
\begin{eqnarray}
\Delta A_{g^\prime}&=&A_{g^\prime,\text{C89}}-A_{g^\prime,\text{S11}}\label{eq:deltaAg} \\
\Delta A_{z^\prime}&=&A_{z^\prime,\text{C89}}-A_{z^\prime,\text{S11}}.  \label{eq:deltaAz} 
\end{eqnarray}
We find, as expected, that the reddening corrections and uncertainties are much larger in the $g^\prime$-band than in 
the $z^\prime$-band and that both reddening estimates generally compare well for $A_{g^\prime}\lesssim1$ and $A_{z^\prime}\lesssim 0.4$. Higher reddenings are more unreliable, probably reflecting the patchiness of the extinction
and irregular distribution of dust clouds. For NGC~6144, NGC~6256, NGC~6544 and NGC~6553 we find $|\Delta A_g|>1$ 
and $|\Delta A_z|>0.4$. 

To further scrutinize the reddening issue, we introduce another new parameter: 
\begin{equation} \label{eq:deltaAgz}
\Delta A_{g^\prime-z^\prime} = (A_{g^\prime}-A_{z^\prime})_{\text{C89}} - (A_{g^\prime}-A_{z^\prime})_{\text{S11}}.
\end{equation}

In practice, $\Delta A_{g^\prime-z^\prime}<0$ results in $g^\prime-z^\prime$ colours which are redder if we would use 
the \cite{Cardelli1989} reddening law instead of the updated maps by \cite{Schlafly2011} to correct for the reddening.

\begin{figure*}
\centering 
\includegraphics[scale=0.87,trim= 3cm 13cm 8cm 6cm]
{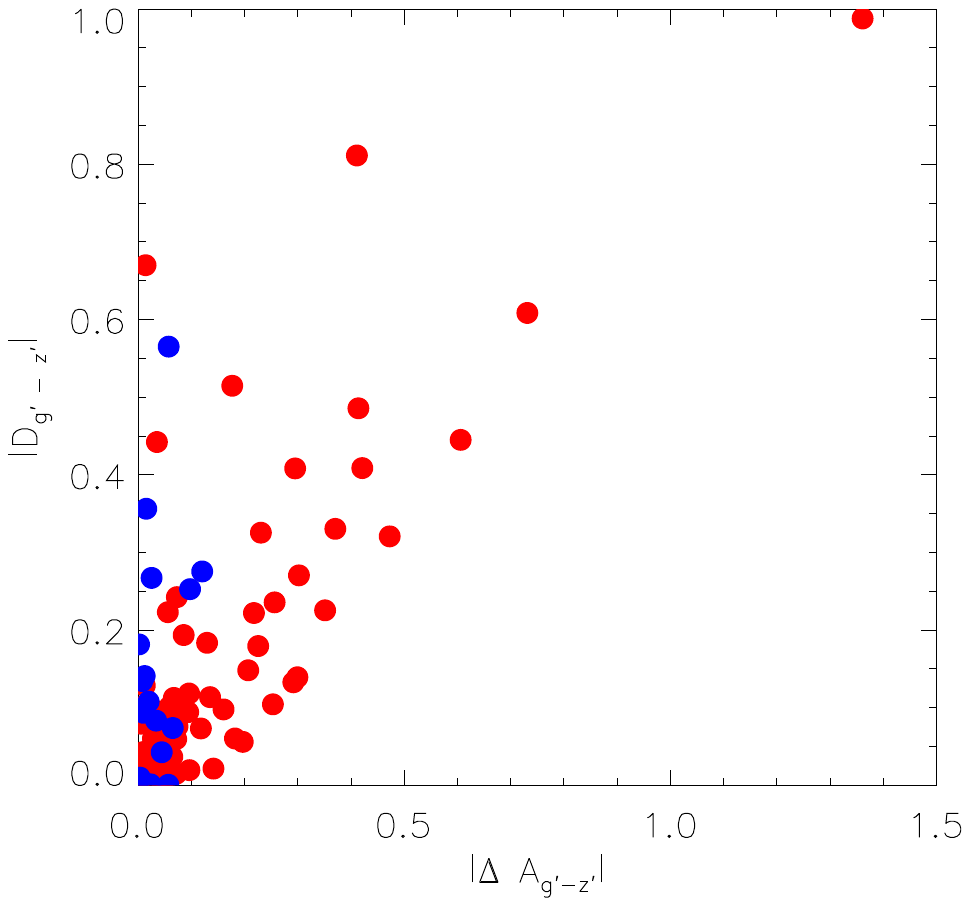}\includegraphics[scale=0.87,trim=
3cm 13cm 8cm 6cm] {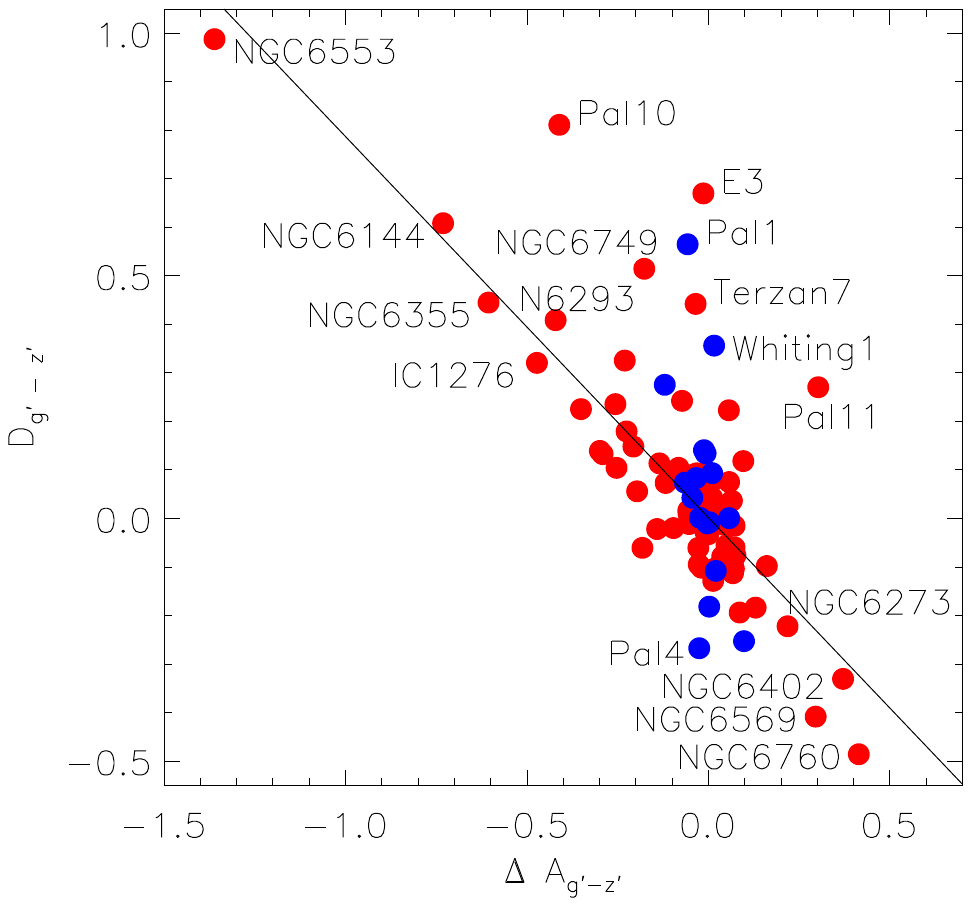}
\caption{Left panel: Absolute scatter around the CMR vs. the absolute reddening coefficient
  difference as defined in Eq.~\ref{eq:deltaAgz}. It is clear that the
  uncertainty in the reddening estimate scales with the distance to
  the CMR. --- Right
  panel: Scatter around the CMR vs. the reddening coefficient
  difference. The black line is a robust fit to the data, as given by
  Eq.~\ref{eq:DeltaAgz_D}. Legend as in Fig.~\ref{GalgzCMR}. Some particular clusters are indicated: these include Pal~1, E~3, the GCs associated with the Sagittarius system (Whiting~1, Terzan~7) and some GCs suffering severe differential reddening (NGC~6144, NGC~6273, NGC~6355, NGC~6402, NGC~6553). See text for more details.  }
\label{fig:DvsDeltaAgz}
\end{figure*}

In the left panel of Fig.~\ref{fig:DvsDeltaAgz} we compare the absolute scatter in the CMR as a function of 
the absolute value of this new parameter. The clusters with  low $ |\Delta A_{g^\prime-z^\prime}| $ and large $ |D_{g^\prime-z^\prime}| $ are 
Pal~1, Whiting~1, Terzan~7 and E~3. 
The origin of these large offsets was discussed in more detail in Section~\ref{sec:cmroutliers}. 
  
The right panel of Fig.~\ref{fig:DvsDeltaAgz} presents the scatter about the CMR as a function of $\Delta
A_{g^\prime-z^\prime}$. Surprisingly, $D_{g^\prime-z^\prime}$ correlates with $\Delta A_{g^\prime-z^\prime}$. 
This is unexpected: in fact it predics that, when:
\begin{equation*}
(A_{g^\prime}-A_{z^\prime})_{\text{C89}}<(A_{g^\prime}-A_{z^\prime})_{\text{S11}},
\end{equation*}
the $g^\prime-z^\prime$ colour (based on the \citealt{Schlafly2011} extinction coefficients) is too
blue and results in  $D_{g^\prime-z^\prime}>0$. This demonstrates that the scatter around the CMR 
and the reliability of the extinction estimate are intimately related. Moreover, GCs located at Galactic latitude $|b\lesssim5|$ (including Pal~10, NGC~6553, NGC~6355 and NGC6760) are known to have unreliable extinction estimates \citep{Schlafly2011}.

Again, E~3, Pal~1, Terzan~7 and Whiting~1 do not follow the general trend. Nevertheless, some other clusters also do not follow the relation either: 
Pal~10, Pal~11 and NGC~6749 all suffer high extinction ($E(B-V)=1.66$, 0.35 and 1.50, respectively). For these clusters, 
it is not the difference between the reddening estimates of \cite{Cardelli1989} and \cite{Schlafly2011} that causes the 
offset in the CMR. However, as these are high reddening clusters, it does illustrate again that the scatter in the CMR 
scales with the reddening.  

We made a robust fit of $D_{g^\prime-z^\prime}$ as a function of $\Delta A_{g^\prime-z^\prime}$ and obtain 
\begin{equation} \label{eq:DeltaAgz_D}
D_{g^\prime-z^\prime} = -0.78\times \Delta A_{g^\prime-z^\prime}- 0.02,
\end{equation}
plotted as a black line in Fig.~\ref{fig:DvsDeltaAgz}. We can use this relation to correct the $g^\prime-z^\prime$
colours: 
\begin{equation*} 
(g^\prime-z^\prime)^*=(g^\prime-z^\prime)+(-0.78\times \Delta
A_{g^\prime-z^\prime} - 0.02)
\end{equation*}
or 
\begin{multline}\label{eq:gzstar}
(g^\prime-z^\prime)^*= \hat{g}-\hat{z}-0.78\times (A_{g^\prime}-A_{z^\prime})_{\text{C89}}\\-0.22\times (A_{g^\prime}-A_{z^\prime})_{\text{S11}}-0.02,
\end{multline}
with $\hat{g}$ and $\hat{z}$ the calibrated but not reddening corrected magnitudes.

\begin{figure}
\centering 
\includegraphics[scale=0.87,trim= 3cm 13cm 8cm 6cm]
{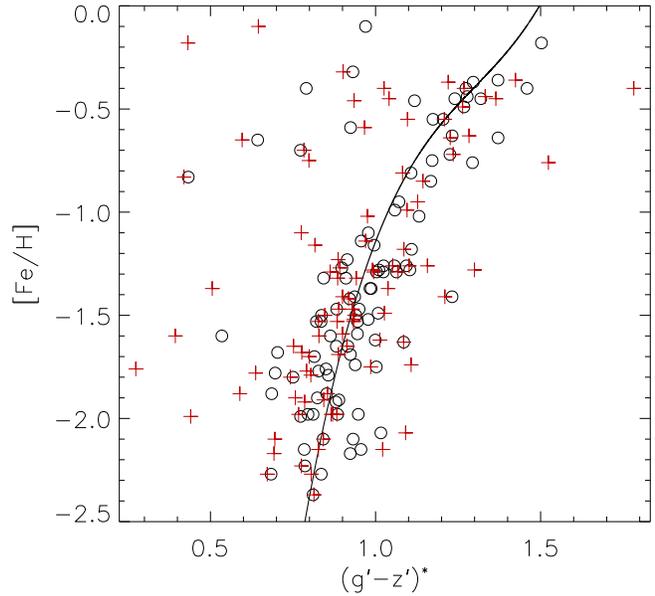}
\caption{[Fe/H] as a function of $(g^\prime-z^\prime)^*$, a colour
  corrected for the reddening uncertainty (defined in
  Eq.~\ref{eq:gzstar}). Corrected colours are indicated with black
  circles. As a reference, red crosses represent the
  $(g^\prime-z^\prime)$, which were also given in Fig.~\ref{GalgzCMR}. The
  solid line presents the CMR as given by Eq.~\ref{eq:gzPython}. }
\label{fig:gz*}
\end{figure}

Fig.~\ref{fig:gz*} presents a CMR for $(g^\prime-z^\prime)^*$, which are represented as black circles. For reference, the
$g^\prime-z^\prime$ colours are given as red crosses. Note the huge difference for e.g. NGC~6553, a high-metallicity 
cluster with a colour correction of more than 1 magnitude. It is clear that the scatter is significantly reduced: the
horizontal RMS for the $g^\prime -z^\prime$ colour is 0.24, while it is 0.16 for the corrected $(g^\prime-z^\prime)^*$
colour. This suggests that a combination of the C89 and S11 reddening laws results in a better extinction estimate. Moreover, the horizontal RMS for the $g^\prime -z^\prime$ colour is also 0.16 when we limit the sample to GCs with $E(B-V)<0.35$. Nevertheless, not all GCs are 
moved towards the CMR by applying the reddening correction of Eq.~\ref{eq:gzstar}. Pal~11 is a high-reddening 10.4 Gyr
 old GC \citep{Lewis2006} that scatters off the CMR when applying this correction. E~3, Pal~1, Terzan~7 and Whiting~1 are not affected by the reddening correction. Again, their 
position in the figure suggests another evolutionary history and younger ages. 

Pal~10, another high reddening GC ($E(B-V)\sim1.66$) in the Sagittarius constellation, is moved towards but not on the CMR by the reddening correction. It is notable that the difference between the corrected $(g^\prime-z^\prime)^*$ colour and the CMR is similar to the difference for Terzan~7 and Whiting~1, which are associated to the Sgr dSph.  

\begin{figure}
\centering 
\includegraphics[scale=0.87,trim= 3cm 13cm 8cm 6cm]
{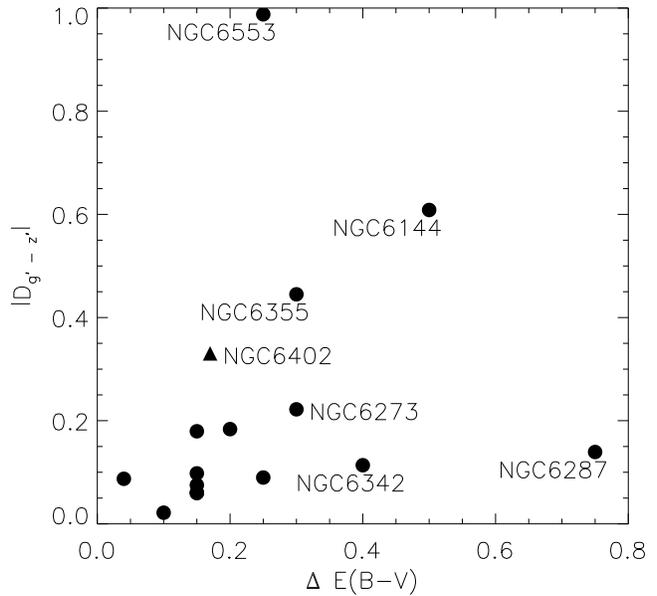}
\caption{Absolute colour difference $|D_{g^\prime - z^\prime}|$ (as defined in Eq.~\ref{eq:D}) as a function of the differential reddening $\Delta E(B-V)$ \citep{AlonsoGarcia2012,ContrerasPena2013}. Some of the clusters with $\Delta E(B-V)>0.15$ show large offsets from the CMR, while clusters with relatively low differential reddening ($\Delta E(B-V)<0.15$) are all close to the CMR.  }
\label{fig:D_deltaEBV}
\end{figure}

Some GCs suffer from substantial differential reddening \citep{Heitsch1999,AlonsoGarcia2012}. To estimate the contribution of the differential reddening
to the photometric error and to the scatter in the CMR, we present in Fig.~\ref{fig:D_deltaEBV} the absolute distance to the CMR as a function of the differential reddening 
$\Delta E(B-V)$ (obtained from \citealt{AlonsoGarcia2012}, supplemented with data from \citealt{ContrerasPena2013} for NGC~6402). Some of the clusters with $\Delta E(B-V)>0.15$ show large 
offsets from the CMR, while clusters with relatively low differential reddening ($\Delta E(B-V)<0.15$) are all close to the CMR. NGC~6553, NGC~6144 
and NGC~6355 are clusters with $\Delta A_{g^\prime-z^\prime}>0.5$. NGC~6287 has $\Delta A_{g^\prime-z^\prime}=-0.3$, but lies remarkably 
close to the CMR. The large $\Delta A_{g^\prime-z^\prime}$ values could suggest that the differential reddening is affecting  the reddening estimate 
for the entire cluster. Nevertheless, NGC6402, NGC~6273, NGC~6553, NGC~6144 and NGC~6355 are right on the relation presented in the right panel of 
Fig.~\ref{fig:DvsDeltaAgz}, suggesting that combining both \cite{Cardelli1989} and \cite{Schlafly2011} reddening laws could resolve the issue. 

In a future paper, we will determine the reddening by fitting isochrones to the colour-magnitude diagrams 
and further discuss this issue. 

\subsection{The Effects of Horizontal Branch morphology}
\begin{figure}
\centering 
\includegraphics[scale=0.85,trim= 2.8cm 13.2cm 9cm 6cm] {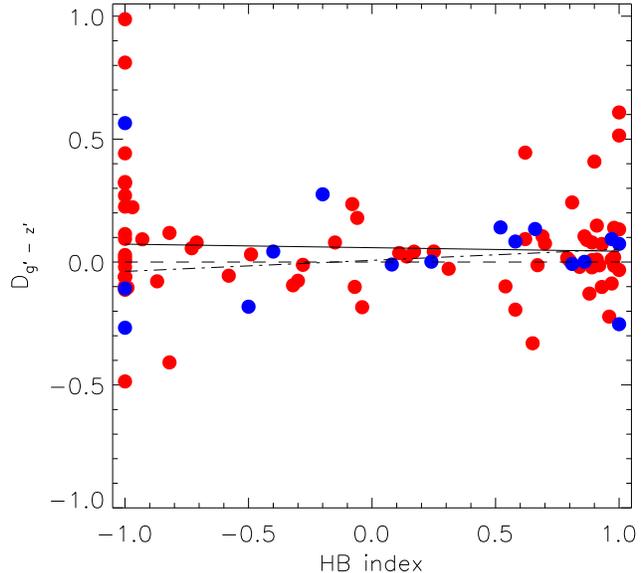}
\caption{Colour difference $D_{g^\prime - z^\prime}$ (as defined in Eq.~\ref{eq:D}) as a
  function of HB index  \citep{Mackey2005}. The best fit is given as a
  solid line, the dashed line indicates the $D_{g^\prime - z^\prime}=0$. The dash-dot line is the best fit for clusters with HB index between $-0.9$ and 0.9. Legend as in Fig.~\ref{GalgzCMR}. See text for more details.}
\label{distance_function_HBindex}
\end{figure}

\cite{Yoon2006} studied the influence of HB morphology on GC colours and indicated that HB stars are the
main drivers behind the non-linearity of the CMR. The CMR presented in Eq.~\ref{eq:gzPython} is clearly non-linear and should account for the influence of the HB stars. However, to check whether the HB morphology contributes to the scatter around the 
CMR we plot in Fig.~\ref{distance_function_HBindex} the colour difference $D_{g^\prime - z^\prime}$ 
as function of the HB morphology index ($\frac{B-R}{B+V+R}$, \citealt{Lee1990,Lee1994}) of \cite{Mackey2005} for 78 GCs. 
The best fit relation 
\begin{equation}
D_{g^\prime - z^\prime}=(0.058\pm 0.023)-(0.014\pm 0.029) \times \frac{B-R}{B+V+R}
\end{equation}
is not statistically significant and is given by the solid line in Fig.~\ref{distance_function_HBindex}. The HB index becomes
insensitive to the HB morphology for very blue and very red HBs \citep[][and references therein]{Catelan2001}, 
which are the ranges which are best populated in our GC sample. Motivated by this argument, we restrict the HB
index range to $[-0.9,0.9]$ to fit the data and find 
\begin{equation}
D_{g^\prime - z^\prime}=(0.01\pm 0.02)+(0.05 \pm 0.04)\times \frac{B-R}{B+V+R},
\end{equation}
presented by the dash-dot line in the same figure. Again, this relation is statistically not significant, which indicates that the non-linear CMR does account well for the influence of the HB morphology on the cluster colour.

To further investigate the effects of the HB morphology we list in Table~\ref{tab:2ndpar} some clusters 
with similar metallicities but different HB structure, the so-called 'second parameter objects'  (e.g.
\citealt{Catelan2001,Caloi2005}). For NGC~288 and NGC~362, one of the best studied 'second parameter 
pairs' of GCs, we find a blue (red, respectively) offset from the CMR, as could be expected from the HB morphology
of these clusters. The same holds for NGC~6205 and NGC~7006, while this is not the case for NGC~5272. In 
most cases $|D_{g^\prime - z^\prime}|$ is smaller than the colour uncertainties; therefore, we conclude that the effect of the HB morphology on the integrated colour is reasonably well accounted for by the non-linear CMR. 

\begin{table}
\centering
\caption{\label{tab:2ndpar} HB morphology (taken from \citealt{Mackey2005}), [Fe/H] and distance to the CMR (as defined in Eq.~\ref{eq:D}) for some crucial clusters regarding the second parameter problem. }
\begin{tabular}{lccc}
\hline
 & HB index & [Fe/H] & $D_{g^\prime - z^\prime}$ \\
\hline
NGC~288 & 0.98 & $-1.32$ & 0.019\\
NGC~362 & $-0.87$ &$-1.26$ & $-0.078$\\
\hline 
NGC 5272 (M 3) & 0.08 & $-1.50$ & $-0.010$\\
NGC 6205 (M 13) & 0.97 & $-1.53$&0.094\\
NGC 7006 & $-0.28$ & $-1.52$&$-0.011$\\
\hline
\end{tabular} \\
\end{table}

\subsection{Age as possible cause for the scatter in CMR}\label{sec:age}
In Section~\ref{sec:cmroutliers}, discussing the young GCs associated to the Sagittarius and Canis Major system, 
we already considered age as possible contributor to the scatter in the CMR. In this section, we will discuss this
issue in some more detail, concentrating on both high-reddening and low-reddening clusters. Recently,
\cite{Forbes2010} made a compilation of the most reliable ages published to date (based on results of
\citealt{Salaris1998, Bellazzini2002, Catelan2002, deAngeli2005,Carraro2007,MarinFranch2009}). We add new 
ages for IC4499, Pal 15 and NGC7006 from \cite{Dotter2011}, NGC6293 \citep{Lee2006}, NGC6402 
\citep{Paust2011}, NGC6553 \citep{Ortolani1995}, Pal 11 \citep{Lewis2006} and Pal 13 \citep{Trouille2002}. Note that \cite{Vandenberg2013} do not give an age estimate for Pal~1, Terzan~7 and E~3, because of the poor quality of the available CMDs.

\begin{figure}
\centering 
\includegraphics[scale=0.85,trim= 2.8cm 13.2cm 9cm 6cm] {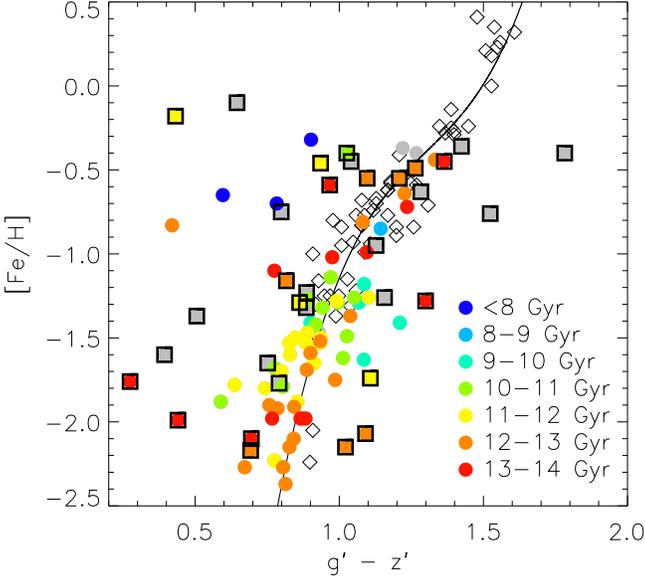}
\caption{[Fe/H] as a function of $g^\prime -z^\prime$, colour-coded by the ages of the GCs. Grey symbols are used for the clusters without age estimate, open diamonds are the M~49 and M~87 GCs. Black boxes indicate high-reddening GCs. As a reference, the CMR (given by Eq.~\ref{eq:gzPython}) is presented by the black solid line. See text for more details. }
\label{fig:age}
\end{figure}

Fig.~\ref{fig:age} presents [Fe/H] as a function of the $g^\prime -z^\prime$ colour, with the clusters being colour-coded depending on their age. Three young low-reddening GCs (Terzan~7, Pal~1 and Whiting~1) were associated with the Canis Major and Sagittarius systems. For the old GC E~3, we discussed in Section \ref{sec:cmroutliers} the photometric uncertainties for the observations of this faint cluster, which suffers significant foreground extinction ($E(B-V)\sim0.3$).  

To pinpoint the influence of age on the $g-z$ colour for the younger GCs, we integrated PARSEC (v1.1) isochrones from \cite{Bressan2012} with a \cite{Kroupa1998} IMF corrected for binaries\footnote{http://stev.oapd.inaf.it/cgi-bin/cmd}. For Pal~1 and Whiting~1 (with [Fe/H]$\sim-0.7$ and an age about 7 Gyr), we used a metallicity $Z\sim0.004$ and obtained SSP integrated $g-z\sim1.1$, much redder than the observed $g-z\sim0.6$ and $\sim0.8$ (respectively). For Terzan~7 (with [Fe/H]$\sim-0.32$ and age of 7.3 Gyr) we adopted $Z\sim0.008$. For this combination, the models predict $g-z\sim1.2$, much redder than $g-z\sim0.9$ based on our observations. The models confirm that the colours of GCs, older than a few Gyrs, are totally dominated by the RGB, which is populous and bright. At old-enough ages it is only the metal abundance that sets the colour of the RGB, suggesting that Pal~1, Whiting~1 and Terzan~7 are peculiar clusters with a different chemical enrichment history. This will be further discussed in Section~\ref{sec:chemistry}.    

It is clear that age is not the main contributor to the scatter in the CMR. Several old high-reddening clusters are offset of the relation (e.g. NGC~6144, NGC~6293, NGC~6402). Pal~15 and NGC~6426 are distant, high-reddening GC (with signs of differential reddening) and are coeval with all other metal-poor GCs, with ages estimated about 13 Gyrs \citep{Dotter2011}. The metal-rich outlier with an age of 12 Gyr is NGC6553, suffering significant differential reddening, as was described in Section~\ref{sec:reddening}.

\subsection{Do mass function variations play a role? } \label{sec:IMF}
In Section~\ref{sec:age}, we quietly assumed that the GC initial mass function (IMF) is well represented by a \cite{Kroupa1998} IMF. Although the IMF of the Milky Way field stars is mostly consistent with \cite{Salpeter1955} or Kroupa IMFs \citep{Bochanski2010}, recent studies discuss the possibility of IMF variations (e.g. \citealt{Conroy2012} and references therein). \cite{Paust2010} suggested that the observed variations of GC present-day mass functions (MF) are related to dynamical evolution, while \cite{Marks2012} claim IMF correlations with cluster density and metallicity. 

Recently, \cite{Hamren2013} gathered the MF slopes published to date (based on results of \citealt{Paust2010,Rosenberg1998,Bellazzini2012, Frank2012, Pulone2003, Dotter2008,Cote1991,Grillmair2001,Jordi2009,DeMarchi2007,Capaccioli1991,Grabhorn1991,Paust2009,Saviane1998,Milone2012}). These authors used a single-sloped power law MF in the form $dN/dm\sim m^{-(1+\alpha)}$ (implying $\alpha=+1.35$ for the classic \cite{Salpeter1955} MF) and cover masses $M\lesssim0.8M_\odot$ of the present day mass function, which are the stellar masses remaining in old clusters. 

\begin{figure}
\centering 
\includegraphics[scale=0.85,trim= 2.8cm 13.2cm 9cm 6cm] {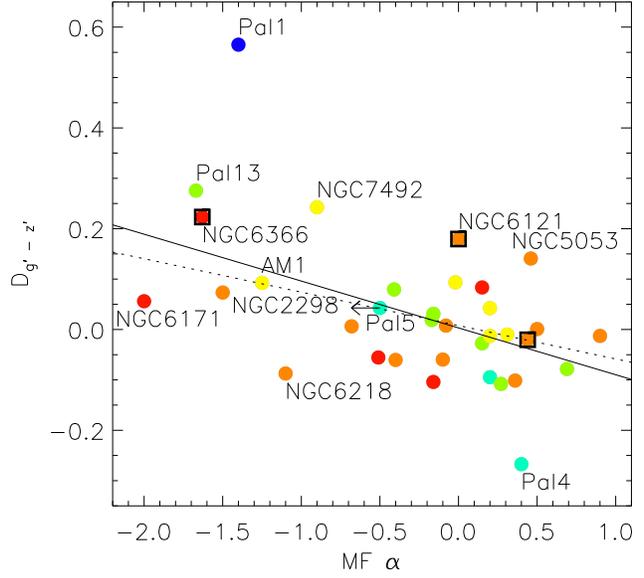}
\caption{Colour difference $D_{g^\prime - z^\prime}$ (as defined in Eq.~\ref{eq:D}) as a
  function of the GC MF slope $\alpha$. Legend as in Fig.~\ref{fig:age}. The solid line, given by Eq.~\ref{eq:IMFall} represents the best robust fit to all the data. The dotted line, given by Eq.~\ref{eq:IMFexclPal1_4}, shows the best robust fit excluding Pal~1 and Pal~4. The slope of the MF of Pal~5 is only an upper limit. See text for more details. }
\label{fig:IMF}
\end{figure}

In Fig.~\ref{fig:IMF} we plot the colour difference $D_{g^\prime - z^\prime}$ (defined in Eq.~\ref{eq:D}) as a function of the MF power law slope for 34 GCs. The clusters are colour-coded with age as in Fig.~\ref{fig:age}. Pal~5 is indicated with an arrow, because only an upper limit for the MF slope was given. Making a robust fit to the data results in 
\begin{equation} \label{eq:IMFall}
D_{g^\prime - z^\prime} = 0.00\pm0.01 - (0.09\pm0.03)\times \alpha,
\end{equation}
which is represented by the solid line in Fig.~\ref{fig:IMF}. Limiting the fit to the low-reddening ($E(B-V)<0.35$) clusters does not alter the fit and using the extinction corrected $(g^\prime-z^\prime)^*$ colours (defined in Section~\ref{sec:reddening}) confirms the correlation. 

Despite the significant scatter among the relation, Pal~1 and Pal~4 are the strongest outliers. Both clusters are faint, resulting in $\sigma_{g-z}\sim0.05$ for Pal~1 and $\sigma_{g-z}\sim0.12$ for Pal~4. Remark that the magnitudes obtained for Pal~1 and Pal~4 are based on SDSS data. Therefore, their colours did not suffer from the complications regarding the CTIO sky determination (as discussed in Paper I). Excluding these clusters to make a robust fit we find 
\begin{equation} \label{eq:IMFexclPal1_4}
D_{g^\prime - z^\prime} = 0.01\pm0.01 - (0.07\pm0.02)\times \alpha, 
\end{equation}
 fully consistent with Eq.~\ref{eq:IMFall}. Nevertheless, if we use the $(g^\prime-z^\prime)^*$ colours excluding Pal~1 and Pal~4, the slope of the relation does become more shallow and is only different from zero at the 1.6 $\sigma$ level. 

The MF slope for Whiting~1 is not given by \cite{Hamren2013}. Nevertheless, \cite{Carraro2007} found that the luminosity function of Whiting~1 is remarkably flat and suggest that the cluster has experienced tidal stripping by the Milky Way. If the mass function is confirmed to be approximately flat, this cluster would be located close to Pal~1 in Fig.~\ref{fig:IMF} and would follow the general trend given by Eq.~\ref{eq:IMFall}: GCs with more bottom-light MFs show a blue offset to the CMR. 

Other GCs showing evidence for tidal stripping by the Milky Way include Pal~5 \citep{Koch2004} and Pal~13 \citep{Cote2002}. Remark that both clusters follow the general trend (given by Eq.~\ref{eq:IMFall}), although the slope of the MF for Pal~5 is just an upper limit.

Remark that NGC~5053, a GC likely associated with the Sgr dSph by \cite{Law2010} shows a blue offset to the CMR. Nevertheless, Eq.~\ref{eq:IMFall} predicts a negligible colour difference based on the MF slope found by \cite{Paust2010}.

Although the coefficients in Eqs.~\ref{eq:IMFall} and \ref{eq:IMFexclPal1_4} are significantly different from zero, we are prudent to conclude any correlations between MF variations and the CMR offset are real. These correlations would not imply the MF varies with metallicity, they only suggest the colour offset to the CMR is related to the MF.

\subsection{Influence of structural parameters in the CMR scatter}\label{sec:structpar}
In this section, we study the CMR scatter and its relation to the structural parameters, which relate to the evolutionary history of the GCs.

\begin{figure*}
\centering 
\includegraphics[scale=0.85,trim= 2.8cm 13.2cm 9cm 6cm] {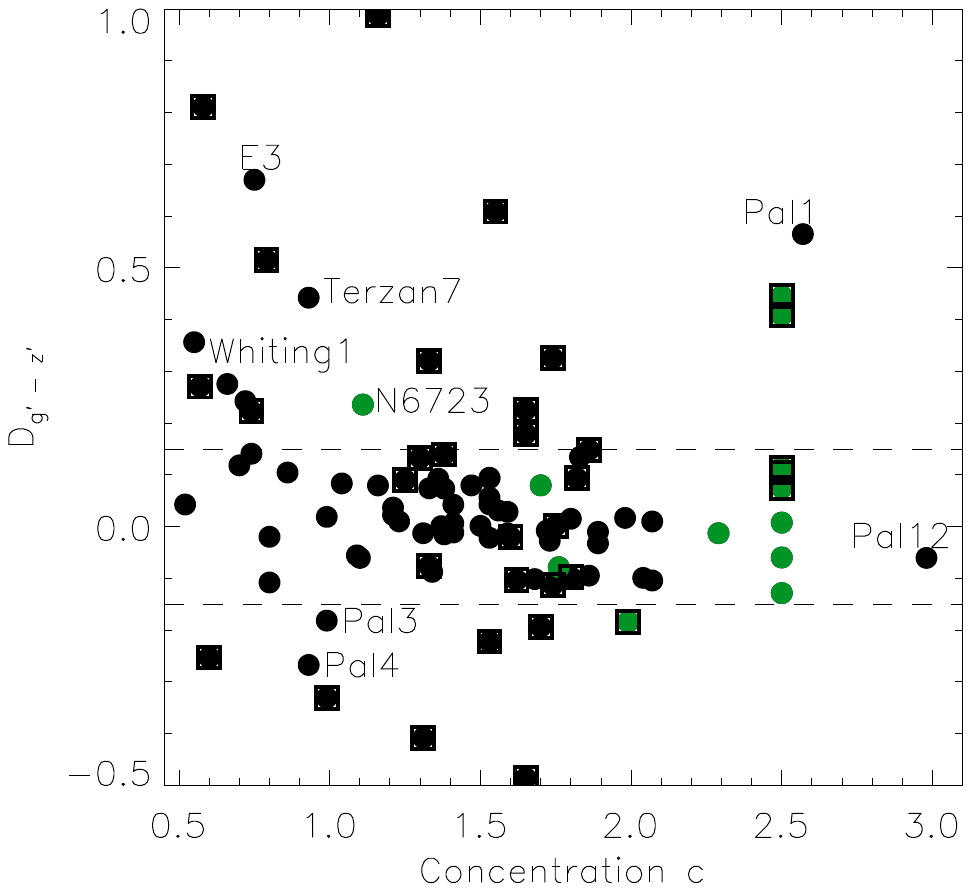}\includegraphics[scale=0.85,trim= 2.8cm 13.2cm 9cm 6cm] {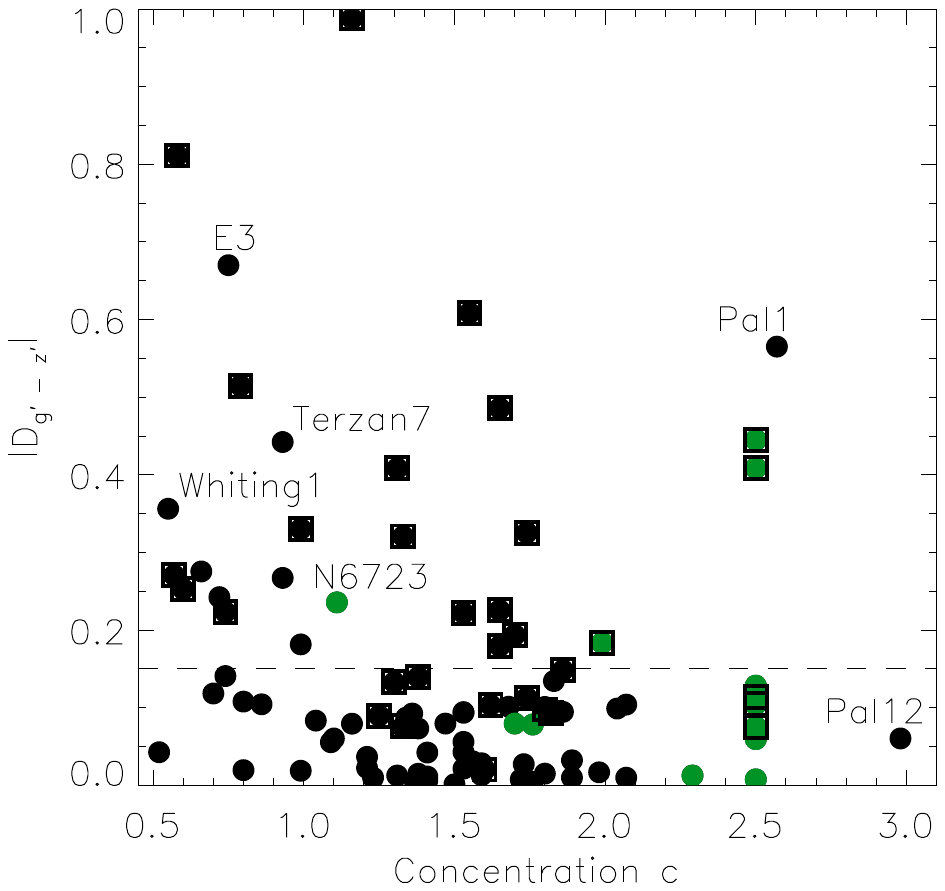}
\
\caption{ Colour difference $D_{g^\prime - z^\prime}$ and absolute colour difference $|D_{g^\prime - z^\prime}|$ (as defined in Eq.~\ref{eq:D}) as a
  function of  the concentration c \citep{Harris1996}. Green circles are used for core-collapsed clusters, black circles for other clusters. Black boxes indicate high-reddened clusters. 
The four low-reddening clusters with low concentration ($c<1$) and $|D_{g^\prime - z^\prime}|>0.15$ that are not indicated in the right panel are Pal~3, Pal~4, Pal~13 and NGC~7492. See text for more details. }
\label{fig:concentration}
\end{figure*}

Fig.~\ref{fig:concentration} presents the colour difference $D_{g^\prime - z^\prime}$ and absolute colour difference $|D_{g^\prime - z^\prime}|$ (as defined in Eq.~\ref{eq:D}) as a function of the concentration c (obtained from the compilation of \citealt{Harris1996} (2010 version), which is based on values taken from \citealt{Trager1993,Trager1995} and \citealt{McLaughlin2005}). In this figure, green symbols are used for core-collapsed clusters. 

It is remarkable that, ignoring Pal~1 and all high-reddening GCs, only low-concentration GCs are offset the CMR. However, all these clusters are faint and we can not exclude that this is the principal cause of the scatter with respect to the CMR.  
Note that NGC~6723, a GC in the Sagittarius constellation, has a similar concentration and metal abundance to Whiting~1 and Terzan~7. However, it has an estimated age of 13.06 Gyr \citep{MarinFranch2009}, hence this GC is much older than Whiting~1 and Terzan~7. 

Pal~1, the GC which might be associated to the Canis Major dSph, has a very different concentration and is the only low-reddening cluster with a concentration $c>1$ that is an outlier in the CMR.

Note that low-reddening core-collapsed clusters (with the exception of NGC~6723) are all close to the CMR. This suggests that the $g^\prime-z^\prime$ colours of GCs are not altered during or after the core-collapse.

\subsection{A Note on the influence of the contamination correction on the scatter
  in the CMR}
In Paper I we described how we used CMDs and
proper motions to
clean out the aperture magnitudes. In this section we check how
effective this correction is and what its influence is on the scatter
in the CMR. 

\begin{figure*}
\centering 
\includegraphics[scale=0.95,trim= 15.5cm 13.3cm 14cm 6cm]{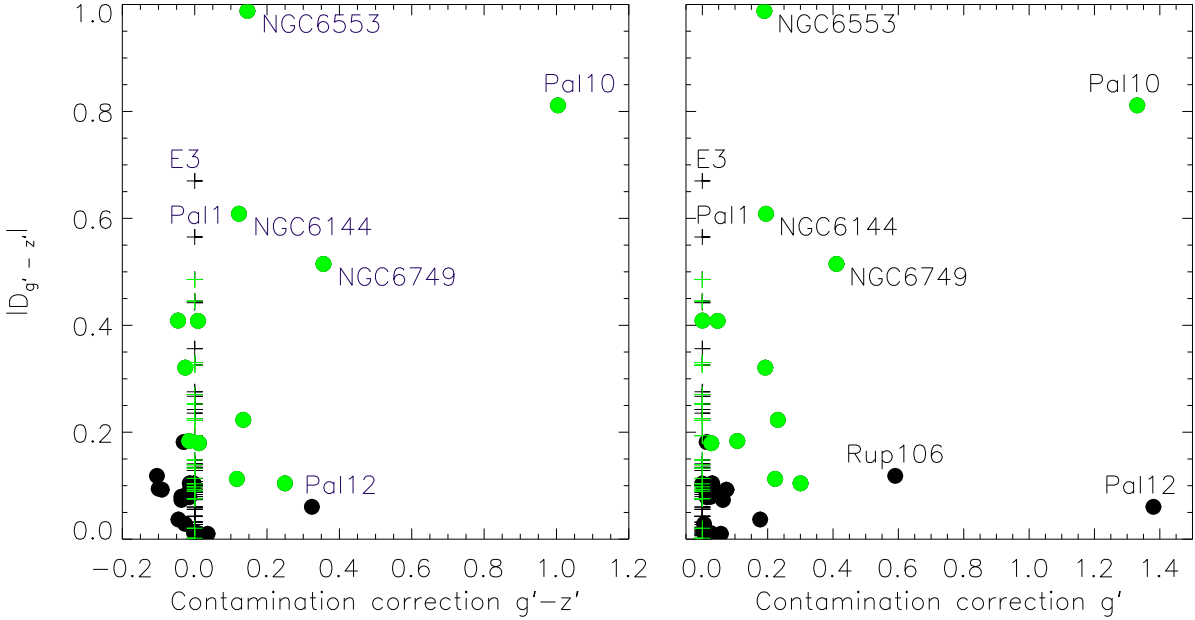}
\caption{Absolute colour difference $|D_{g^\prime - z^\prime}|$ as a
  function of the contamination correction in $g^\prime$ and $g^\prime
  - z^\prime$. Crosses indicate clusters without contamination
  correction based on the CMDs, filled circles represent the GCs for which 
  contamination corrections were applied. Green symbols are used for clusters with $E(B-V)\geq0.35$. }
\label{fig:DvsCMDcorr}
\end{figure*}

Fig. \ref{fig:DvsCMDcorr} presents the absolute colour difference
$|D_{g^\prime - z^\prime}|$ (defined in Eq. \ref{eq:D}) as a function of
the $g^\prime$ magnitude correction and the $(g^\prime  - z^\prime)$ colour
correction. Only three low-reddening clusters deviate strongly from
the CMR. These clusters (E 3, Pal 3 and Pal 13) are very faint which
is reflected in their magnitude errors. It was not possible to
obtain a decent CMD for these clusters so no magnitude correction was applied. 

The need to clean out the contamination, especially for faint
clusters, is illustrated well by Pal~12. For this halo cluster, only one very
bright foreground star was identified in the CMD and removed from the
aperture photometry. This resulted in magnitude corrections of
 1.38 (1.20, 1.15, 1.06) mag. in $g^\prime$ ($r^\prime$, $i^\prime$,
 $z^\prime$), yielding a 0.32 mag. correction for the contamination in
 $g^\prime-z^\prime$. Note that for this cluster $|D_{g^\prime -
   z^\prime}|=0.06$, so the magnitude correction significantly moved
 the GC towards the CMR relation.

Nevertheless, as was discussed in Section \ref{sec:reddening}, the main source of the scatter in the CMR is the 
reddening uncertainty.   

\subsection{GCs with a different chemical evolution? } \label{sec:chemistry}
In Section~\ref{sec:cmroutliers} we discussed the particular position in colour-metallicity space of Whiting~1, Pal~1, E~3 and Terzan~7. In the previous sections, we excluded age and HB morphology as the origin for the offset for these low-reddening clusters. For E~3 and Terzan~7, we can not exclude photometric uncertainties cause the offset. In this section, we reexamine the issue, but also include NGC~6723, NGC~7492 and Pal~13, which are other low-reddening GCs showing similar blue offsets to the [Fe/H] CMR (see Fig.~\ref{EGgzCMR}). 

NGC~6723 is a low-reddening cluster in the Sagittarius constellation, which was also discussed in Section~\ref{sec:structpar}. However, \cite{Forbes2010} listed this cluster among the Milky Way GCs and did not include this cluster in their subsample associated to the Sagittarius dSph. With an age of about 13 Gyr, it is much older than Terzan~7 and Whiting~1. Note that NGC~6723 and NGC~7492 also show a similar offset to the CaT CMR as Terzan~7 (see Fig.~\ref{Wsaviane}). Pal~13 is a sparse distant halo cluster which is about to be devoured by the Milky Way \citep{Siegel2001}.

Only age and chemical composition determine the colours of the RGB stars which dominate the magnitudes of the GCs, hence it is not unreasonable to consider the chemical history as the possible origin for the colour offset. Moreover, \cite{Sakari2011} show that Pal~1, a low surface brightness cluster with a sparse red giant branch, has a very unusual chemistry: the cluster does not show the Na-O anti-correlation and the neutron-capture elements show different abundances than for standard Galactic GCs. 

Whiting~1 and Terzan~7 are two GCs associated with the Sgr dSph that are offset the CMR, the latter cluster being formed during the main episode of star formation in the Sgr system \citep{Bellazzini1999}. \cite{SmeckerHane2002} show that not only the GCs associated to the Sgr system can be peculiar: the red giant stars in the Sgr dSph galaxy span a wide range of metallicities and show very unusual abundance variations (both for $\alpha$ abundances as for neutron-capture elements), inferring an extended period of star formation and chemical enrichment with considerable mass loss.

\cite{Mackey2004} suggested an extragalactic origin for the old outer halo cluster NGC~7492, a cluster located at $R_{GC}=25 \text{ kpc}$. \cite{Majewski2004} find that it is unlikely that this cluster is a Sagittarius remnant, though these authors do not fully exclude this possibility. Based on a sample of four RGB stars, \cite{Cohen2005} recover the well-known Na-O anti-correlation and find evidence for a chemical history (including neutron capture processes) similar to that of inner halo GCs with similar [Fe/H]. 

Several GCs associated to the Canis Major dSph (NGC~1851, NGC~1904, NGC~2298, NGC~2808, NGC~4590 and Rup~106) are on the CMR.  In our sample, Pal~1 is the only GC related to the Canis Major system \citep{Forbes2010} that is offset the CMR. However, the existence of the Canis Major structure is under debate. It is not clear whether this system is produced by a collision with a dSph satellite galaxy or if it is caused by a warp in the Galactic disk, combined with the spiral arm populations of the Milky Way (see e.g. \citealt{Mateu2009}). 

If the latter turns out to be true, it is even more remarkable that several clusters linked to the Sagittarius stream, which is then the only genuine dSph stream being accreted to the Milky Way, are outliers in the CMR. However, other GCs associated to the Sagittarius system by \cite{Forbes2010} also fall close the CMR (including NGC~6715, Pal~12, NGC~4147 and NGC~5634), thus not all GCs related to dSph galaxies have somehow particular colours. 

\cite{Bellazzini1999} claimed that the stellar content and the star formation history of the Sgr dSph appears very similar to those of other dSph galaxies. Most of the GCs that are consistently off the CMR are related to the Sagittarius dSph in some way. Accurate colours and metallicities for GCs residing in dwarf galaxies are highly desirable to check our findings. If it is true that dwarf galaxies host a number of peculiar GCs, we should be able to find these GCs as well in massive galaxies, which are assembled by accreting such objects in CDM theories. However, these peculiar clusters are faint thus detecting them will be observationally challenging.

Note that not all clusters with particular chemical properties are offset the CMR. 
\cite{Cohen2004} demonstrate the chemical peculiarities of Pal~12, another GC associated to the Sagittarius system \citep{Forbes2010}. Due to the large contamination correction (as shown in Fig.~\ref{fig:DvsCMDcorr}), this cluster falls right on the CMR. NGC~2419, another outer halo cluster with chemical peculiarities \citep{Cohen2012}, is not off the CMR.

\section{Colour bimodality} \label{sec:colourbim}
\begin{figure*}
\centering 
\includegraphics[scale=0.87,trim=2.8cm 13.1cm 9cm 5.7cm] {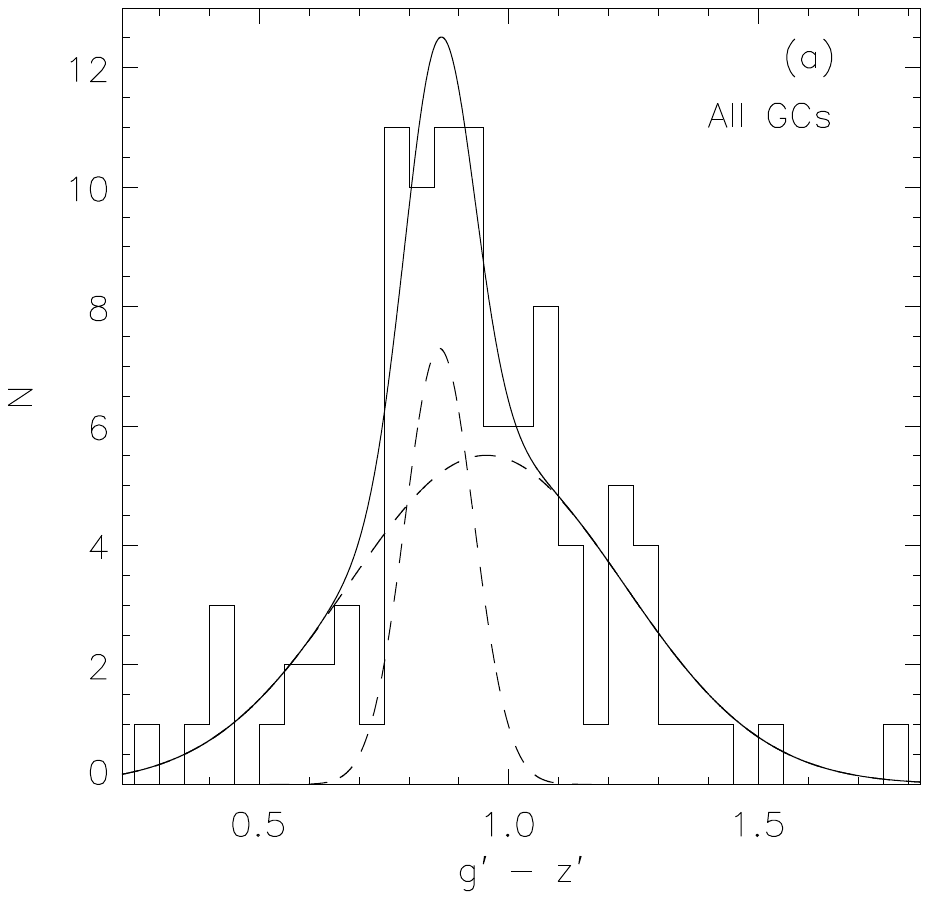}\includegraphics[scale=0.87,trim=2.8cm 13.1cm 9cm 5.7cm] {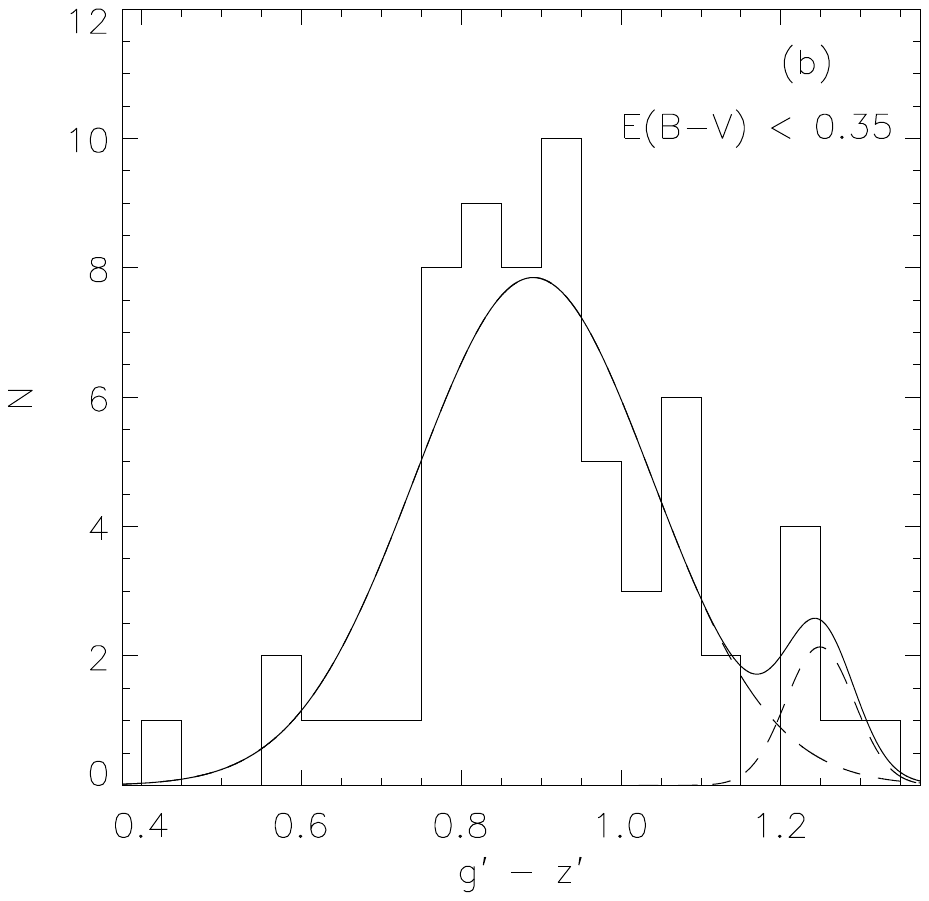}
\caption{$(g^\prime - z^\prime)$ colour distribution for different subsamples of the Galactic GCs. The \textsc{gmm} parameters describing the fits of the Gaussian distributions are tabulated in Table~\ref{tab:gmm}. The solid line is the sum of the two Gaussians obtained with \textsc{gmm}. In general, the distributions are not strongly bimodal, which is unexpected when bearing in mind the bimodal metallicity distribution of the MW. }
\label{gzbimodality}
\end{figure*}

It is widely known that the metallicity distribution of Galactic GCs is bimodal. If the $g^\prime - z^\prime$ colour is
representative of the metallicity then one would expect to find bimodality in the distribution of $g^\prime - z^\prime$ 
colour. However, we ran a Gaussian mixture modelling algorithm (\textsc{gmm} -- \citealt{MuratovGnedin2010}) on the 
full Galactic $g^\prime - z^\prime$ distribution and did not find evidence for a bimodal distribution (Fig.~\ref{gzbimodality}, panel (a)). 

The \textsc{gmm} parameters for the colour distribution and the corresponding [Fe/H] distribution of the full GC sample are listed in
Table~\ref{tab:gmm}. It is clear that the colour distributions are not 
strictly bimodal and could just be skewed unimodal distributions. It would imply that the bimodal Galactic metallicity
distribution transforms into a skewed unimodal distribution as a result of the non-linear CMR. This is in fact the opposite 
case of a unimodal metallicity distribution transforming into a bimodal colour distribution \citep[which was studied in][]{Yoon2006}. Moreover, about half of the metal-poor clusters are attributed to the metal-rich peak based on their colours. 
Note that more than half of the smaller ACSVCS galaxies from \cite{Peng2006}, hosting a similar number of GCs as our current sample, did not exhibit strong colour bimodality either.

\begin{table*}
\centering
\caption{\label{tab:gmm} The results of the \textsc{gmm} analysis for the
  distributions shown in the different panels of
  Fig.~\ref{gzbimodality}: (a) all GCs, (b) $E(B-V) < 0.35$. Case (c) shows the \textsc{gmm} results for the colour distribution (presented in Fig.~\ref{gztrimodality}) for low-reddening GCs, excluding two Sagittarius GCs (Terzan~7 and Whiting~1), Pal~1 and GCs with $\sigma_{g-z}>0.1$. Cases
  (a)$^*$ and (c)$^*$ present the corresponding [Fe/H] distributions for cases
  (a) and (c) (presented in Fig.~\ref{fig:histFeH}). See text for more details. }
\begin{tabular}{lccccc}
\hline
&(a) & (b) &(c) & (a$^*$)& (c$^*$)\\
\hline
 $\mu_1$ & $0.86\pm 0.12$& $0.89\pm 0.053$ & $0.86\pm0.04$& $-1.59\pm 0.05$&$-1.65\pm0.08$\\
 $\mu_2$ & $0.96\pm 0.27$& $1.25\pm 0.11$& $1.07\pm0.03$& $-0.52\pm 0.05$&$-0.52 \pm0.20$\\
  $\mu_3$ & ... & ... &  $1.25 \pm 0.03$  & ...&...\\
 $\sigma_1$& $0.07\pm 0.07$ & $0.14\pm 0.03$ & $0.10 \pm 0.02$& $0.38\pm 0.04$&$0.38\pm 0.041$ \\
 $\sigma_2$& $0.28\pm0.13$& $0.06\pm 0.03$  & $0.03 \pm 0.01$& $0.19\pm 0.04$&$0.16\pm 0.041$\\
 $\sigma_3$& ... & ...&    $0.04 \pm 0.01$ & ...&...\\
  $N$&   96 &63  & 56 &   96& 56\\
  D & $0.47\pm 1.76$& $3.28\pm 0.94$& $5.11\pm1.49$& $3.59\pm 0.36$&$3.67\pm0.81$\\
  $p(\chi^2)$& 0.06& 0.45& 0.23& 0.001&0.055\\
  $p(DD)$& 0.88& 0.21& 0.29& 0.11&0.003\\
  $p(kurt)$& 0.98& 0.87& 0.66&0.01&0.583\\
  Ratio &26 : 74& 93 : 7& 80 : 11 : 9&  75 : 25&83:17\\
  $kurt$ &1.14&0.34&$-0.09$& $-0.97$&$-0.18$\\
\hline
\end{tabular} 
\end{table*}

As our sample is not large (compared to massive galaxies with extensive GC systems) and there are some outliers in the 
CMR, with colour determinations affected by high reddening, we decided to analyse a subsample limited to clusters with 
low reddening. The distribution for this subset is presented in panel (b) of Fig.~\ref{gzbimodality}. In spite of the imposed constraints, no colour bimodality is found. The corresponding \textsc{gmm} parameters are listed in
Table~\ref{tab:gmm}.

\begin{figure}
\centering 
\includegraphics[scale=0.87,trim=2.8cm 13.1cm 9cm 5.7cm] {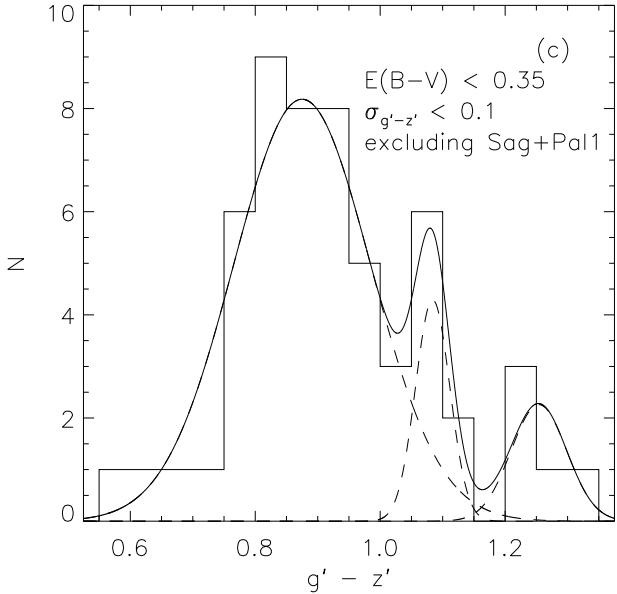}
\caption{Trimodal $(g^\prime - z^\prime)$ colour distribution for a Galactic GC low-reddening subsample with small colour errors and excluding Whiting~1, Terzan~7 and Pal~1. The \textsc{gmm} parameters describing the fits of the Gaussian distributions are tabulated in Table~\ref{tab:gmm}. The solid line is the sum of the Gaussians obtained with \textsc{gmm}. See text for more details. }
\label{gztrimodality}
\end{figure}

It is known \textsc{gmm} is susceptible to outliers, especially long tails  \citep{MuratovGnedin2010,Blakeslee2012}. In a final attempt to recover the $g^\prime - z^\prime$ Galactic colour distribution, we make a new subset of Galactic GCs, excluding GCs with a colour error larger than 0.1 mag, two Sagittarius CMR outliers (Terzan~7 and Whiting~1) and Pal~1 (for reasons described above).
Fig.~\ref{gztrimodality} presents the $g^\prime - z^\prime$ colour distribution for this subset, which is trimodal ($D>2$ and $kurt<0$, although the $p$ value suggests that it is not very statistically significant). \textsc{gmm} parameters are listed in Table~\ref{tab:gmm}. 

The clusters in the reddest peak of the histogram (NGC~104, NGC~6356, NGC~6352, NGC~6624 and Pal~8) are all bulge or thick disk clusters. \cite{Heasley2000} proposed NGC~104 (47~Tuc), NGC~6352, NGC~6624 had a common origin, but \cite{Gao2007} did not assign the GCs to the same accretion streams. \cite{Vandenberg2013} found a bifurcation in their age-metallicity diagram of clusters with disk-like kinematics. It is remarkable that NGC~104, NGC~6352 and NGC~6624 all pertain to the second branch in their diagram. No age estimate was obtained in the latter study for NGC~6356 and Pal~8, which are subject to considerable reddening ($E(B-V)\sim0.3$). 

\cite{Peng2006} found VCC~798 (NGC~4382/M~85) as the best candidate for a trimodal colour distribution. This galaxy is classified as $T=-1$ in the RC3 \citep{Corwin1994}, indicating pure S0, and has a very strong disk component, although the inclination angle makes it appear less obvious. It is tempting to speculate that the trimodal colour distribution is linked with the disk component. Nevertheless, the giant elliptical galaxy NGC~4365 also hosts three subpopulations of GCs \citep{Brodie2005,Blom2012a,Blom2012b}, making a link with the disk less probable. For M~31, a spiral galaxy, some evidence for a trimodal distribution was found, although it is not completely clear how many subpopulations are present \citep{Perrett2002}.

\begin{figure}
\centering 
\includegraphics[scale=0.87,trim=2.8cm 13.1cm 9cm 5.7cm] {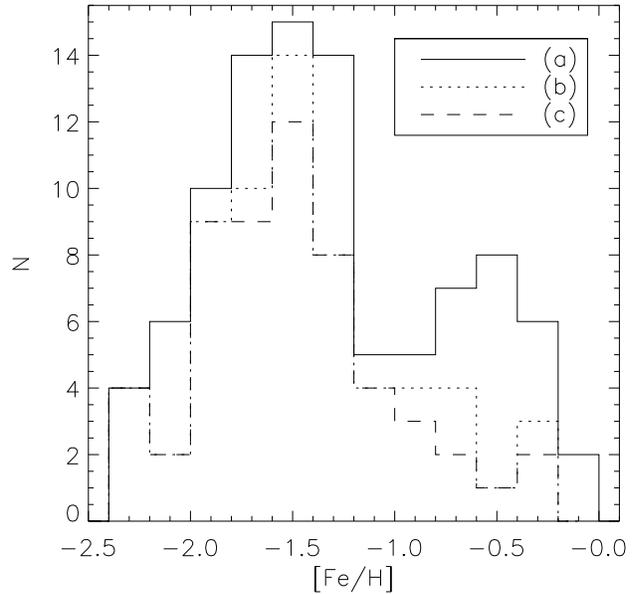}
\caption{[Fe/H] distribution of the different subsamples. Case (a) presents all GCs, case (b) is limited to low-reddening clusters with $E(B-V)<0.35$  and case (c) is limited to low-reddening clusters with small colour errors and excludes GCs associated to Sagittarius. It is clear that the strong [Fe/H] bimodality disappears when limiting the sample to low-reddening clusters.  }
\label{fig:histFeH}
\end{figure}

In Fig.~\ref{fig:histFeH} we present the [Fe/H] distributions of the different subsamples. It is clear that the [Fe/H] distribution for all GCs (case (a)) is bimodal, which is confirmed by the \textsc{gmm} results given in Table~\ref{tab:gmm}. It is clear that the strong [Fe/H] bimodality vanishes limiting the sample to the low-reddening clusters, because a significant fraction of the metal-rich GCs is located towards the bulge of the Galaxy, where the reddening is significant. Note that the \textsc{gmm} parameters for the [Fe/H] distribution associated to case (c) are not conclusive: the peak separation D favours a bimodal distribution, while the second peak is not clearly apparent in Fig.~\ref{fig:histFeH}. The rather large $p$ values favour a unimodal distribution. It was not possible with \textsc{gmm} to fit a trimodal distribution to the case (c) [Fe/H] distribution. Therefore it is rather normal no bimodal colour distributions were found for the different subsamples and it nicely illustrates that selection effects can complicate the correct interpretation of the colour and metallicity distributions. 

\section{Summary} \label{sec:results}
In the current study we used our integrated optical photometry (presented in Paper I) to confirm and improve existing CMRs. For the $(g^\prime -z^\prime)$-[Fe/H] relation 
we double the number of Galactic GCs used in the fit when comparing to the earlier studies by \cite{Peng2006} and \cite{Blakeslee2010}. Moreover, these authors relied on pure aperture photometry and did not correct for foreground
contamination. Nevertheless, we rely on the same extragalactic data to extend the metallicity range. Furthermore, we 
confirm the $(g^\prime-i^\prime)$-[Fe/H] relation of \cite{Sinnott2010}, for the first time with Galactic GCs. However,
we find an offset at both the metal-rich and metal-poor end of their $(g^\prime-z^\prime)$-[Fe/H] CMR. We also demonstrate that the CaT metallicity indicator behaves non-linear when compared to the $(g^\prime-i^\prime)$ and $(g^\prime-z^\prime)$ colours. 

We scrutinise the influence of the reddening estimate on the scatter in the CMR and demonstrate that this scatter can be significantly 
reduced by combining \cite{Cardelli1989} and \cite{Schlafly2011} reddening laws. We also discuss how the scatter 
in the CMR is influenced by the contamination correction, differential reddening, HB morphology, age, present-day mass function variations and structural parameters. 

We find a group of clusters which lie conspicuously off the Galactic CMR: with one possible exception, all these
objects are associated with the Sagittarius dwarf or the proposed Canis Major dwarf. This might imply that a subset of globular
clusters belonging to dwarf spheroidal galaxies are a different population from those found in the Milky Way and other
bright local group members. If so, it will not be possible to build our globular cluster system from mergers of dwarf
galaxies with the early Milky Way, unless such objects are radically different from present-day dwarf spheroidals
in the Local Group.

\section*{Acknowledgments}
We thankfully acknowledge the anonymous referee for very useful and thought-provoking comments. 
We would like to thank Giovanni Carraro, Ivo Saviane, Caroline Foster and Sven De Rijcke for
fruitful discussions, Peter Camps for the implementation of the
fitting routine in Python.  
JV acknowledges the support of ESO through a studentship. JV and MB
acknowledge the support of the Fund for Scientific Research Flanders
(FWO-Vlaanderen). AJ acknowledges support by the Chilean Ministry for
the Economy, Development, and TourismÕs Programa Iniciativa
Cient\'iÞca Milenio through grant P07-021-F, awarded to The Milky Way
Millennium Nucleus.

The authors are grateful to CTIO for the hospitality and the dedicated
assistence during the numerous observing runs. 

This research has made use of NASA's Astrophysics Data System and the
NED which is operated by the Jet Propulsion Laboratory, California
Institute of Technology, under contract with the National Aeronautics
and Space Administration. 

This research made use of Montage, funded by the National Aeronautics and Space Administration's Earth Science Technology Office, Computation Technologies Project, under Cooperative Agreement Number NCC5-626 between NASA and the California Institute of Technology. Montage is maintained by the NASA/IPAC Infrared Science Archive.

For this part of the research, we have made extensive use of the European Virtual
Observatory applications \textsc{aladin} \citep{Bonnarel2000} and \textsc{topcat}
\citep{Taylor2005}. The Virtual Observatory is a project designed to provide the 
astronomical community with the data access and the research tools necessary to
enable the exploration of the digital, multi-wavelength universe resident in the astronomical 
data archives. We used the applications provided by
astrometry.net\footnote{http://astrometry.net/}.

\bibliographystyle{mnras}
\bibliography{references}

\label{lastpage}
\end{document}